\pdfoutput=1
\documentclass[12pt,a4paper]{article}
\usepackage{multirow}
\usepackage{ifthen} % for conditional statements
\newboolean{pdflatex}
\setboolean{pdflatex}{true} % False for eps figures 

\newboolean{articletitles}
\setboolean{articletitles}{true} % False removes titles in references

\newboolean{uprightparticles}
\setboolean{uprightparticles}{false} %True for upright particle symbols

\newboolean{inbibliography}
\setboolean{inbibliography}{false} %True once you enter the bibliography

\def\paperauthors{LHCb collaboration} % Leave as is for PAPER and CONF
\def\paperasciititle{Measurement of the time-integrated CP asymmetry in D/^0 -> K/_S K/_S decays} % Set ASCII title here
\def\papertitle{Measurement of the time-integrated \CP asymmetry in $\Dz \rightarrow \KS\KS$ decays} % Latex formatted title
\def\paperkeywords{{High Energy Physics}, {LHCb}} % Comma separated list
\def\papercopyright{\the\year\ CERN for the benefit of the LHCb collaboration} % new since 9/Apr/2018
\def\paperlicence{CC-BY-4.0 licence}
\def\paperlicenceurl{https://creativecommons.org/licenses/by/4.0/}

\newcommand\Tspc{\rule{0pt}{2.6ex}} 
\newcommand\Bspc{\rule[-1.2ex]{0pt}{0pt}}

\usepackage[top=1in, bottom=1.25in, left=1in, right=1in]{geometry}

\usepackage{microtype}
\usepackage{lineno}  % for line numbering during review
\usepackage{xspace} % To avoid problems with missing or double spaces after
\usepackage{caption} %these three command get the figure and table captions automatically small

\usepackage{graphicx}  % to include figures (can also use other packages)
\usepackage{color}
\usepackage{colortbl}
\graphicspath{{./figs/}} % Make Latex search fig subdir for figures

\usepackage{amsmath} % Adds a large collection of math symbols
\usepackage{amssymb}
\usepackage{amsfonts}
\usepackage{upgreek} % Adds in support for greek letters in roman typeset

\newcommand*\patchAmsMathEnvironmentForLineno[1]{%
\expandafter\let\csname old#1\expandafter\endcsname\csname #1\endcsname
\expandafter\let\csname oldend#1\expandafter\endcsname\csname
end#1\endcsname
 \renewenvironment{#1}%
   {\linenomath\csname old#1\endcsname}%
   {\csname oldend#1\endcsname\endlinenomath}%
}
\newcommand*\patchBothAmsMathEnvironmentsForLineno[1]{%
  \patchAmsMathEnvironmentForLineno{#1}%
  \patchAmsMathEnvironmentForLineno{#1*}%
}
\AtBeginDocument{%
\patchBothAmsMathEnvironmentsForLineno{equation}%
\patchBothAmsMathEnvironmentsForLineno{align}%
\patchBothAmsMathEnvironmentsForLineno{flalign}%
\patchBothAmsMathEnvironmentsForLineno{alignat}%
\patchBothAmsMathEnvironmentsForLineno{gather}%
\patchBothAmsMathEnvironmentsForLineno{multline}%
\patchBothAmsMathEnvironmentsForLineno{eqnarray}%
}

\usepackage{hyperxmp}

\usepackage[pdftex,
            pdfauthor={\paperauthors},
            pdftitle={\paperasciititle},
            pdfkeywords={\paperkeywords},
            pdfcopyright={Copyright (C) \papercopyright},
            pdflicenseurl={\paperlicenceurl}]{hyperref}

\usepackage[all]{hypcap} % Internal hyperlinks to floats.

\usepackage{xspace} 
\def\lhcb {\mbox{LHCb}\xspace}

\def\belle  {\mbox{Belle}\xspace}
\def\cleo   {\mbox{CLEO}\xspace}

\def\MagUp {\mbox{\em Mag\kern -0.05em Up}\xspace}

\ifthenelse{\boolean{uprightparticles}}%
{

 \def\Ppi         {\ensuremath{\uppi}\xspace}

 \def\PDelta      {\ensuremath{\Delta}\xspace}                 
 \def\PXi      {\ensuremath{\Xi}\xspace}                 
 \def\PLambda      {\ensuremath{\Lambda}\xspace}                 
 \def\PSigma      {\ensuremath{\Sigma}\xspace}                 
 \def\POmega      {\ensuremath{\Omega}\xspace}                 
 \def\PUpsilon      {\ensuremath{\Upsilon}\xspace}                 
 
 \def\PB      {\ensuremath{\mathrm{B}}\xspace}                 
                  
 \def\PD      {\ensuremath{\mathrm{D}}\xspace}

 \def\PK      {\ensuremath{\mathrm{K}}\xspace}

 \def\Pb      {\ensuremath{\mathrm{b}}\xspace}                 
 \def\Pc      {\ensuremath{\mathrm{c}}\xspace}

 \def\Pi      {\ensuremath{\mathrm{i}}\xspace}

}
{

 \def\Ppi         {\ensuremath{\pi}\xspace}

 \mathchardef\PDelta="7101
 \mathchardef\PXi="7104
 \mathchardef\PLambda="7103
 \mathchardef\PSigma="7106
 \mathchardef\POmega="710A
 \mathchardef\PUpsilon="7107
                  
 \def\PB      {\ensuremath{B}\xspace}                 
                  
 \def\PD      {\ensuremath{D}\xspace}

 \def\PK      {\ensuremath{K}\xspace}

 \def\Pb      {\ensuremath{b}\xspace}                 
 \def\Pc      {\ensuremath{c}\xspace}

 \def\Pi      {\ensuremath{i}\xspace}

}

\makeatletter
\ifcase \@ptsize \relax% 10pt
  \newcommand{\miniscule}{\@setfontsize\miniscule{4}{5}}% \tiny: 5/6
\or% 11pt
  \newcommand{\miniscule}{\@setfontsize\miniscule{5}{6}}% \tiny: 6/7
\or% 12pt
  \newcommand{\miniscule}{\@setfontsize\miniscule{5}{6}}% \tiny: 6/7
\fi
\makeatother

\DeclareRobustCommand{\optbar}[1]{\shortstack{{\miniscule (\rule[.5ex]{1.25em}{.18mm})}
  \\ [-.7ex] $#1$}}

\def\cquark    {{\ensuremath{\Pc}}\xspace}

\def\bquark    {{\ensuremath{\Pb}}\xspace}

\def\pion   {{\ensuremath{\Ppi}}\xspace}

\def\pip    {{\ensuremath{\pion^+}}\xspace}
\def\pim    {{\ensuremath{\pion^-}}\xspace}

\def\kaon    {{\ensuremath{\PK}}\xspace}
  \def\Kbar    {{\kern 0.2em\overline{\kern -0.2em \PK}{}}\xspace}

\def\KorKbar    {\kern 0.18em\optbar{\kern -0.18em K}{}\xspace}
\def\Kz      {{\ensuremath{\kaon^0}}\xspace}
\def\Kzb     {{\ensuremath{\Kbar{}^0}}\xspace}
\def\Kp      {{\ensuremath{\kaon^+}}\xspace}
\def\Km      {{\ensuremath{\kaon^-}}\xspace}

\def\KS      {{\ensuremath{\kaon^0_{\mathrm{ \scriptscriptstyle S}}}}\xspace}

\newcommand{\KSsub}[1]{{\ensuremath{\kaon^0_{\mathrm{ \scriptscriptstyle S}\mathrm{#1}}}}\xspace}

  \def\Dbar    {{\kern 0.2em\overline{\kern -0.2em \PD}{}}\xspace}
\def\D       {{\ensuremath{\PD}}\xspace}

\def\DorDbar    {\kern 0.18em\optbar{\kern -0.18em D}{}\xspace}
\def\Dz      {{\ensuremath{\D^0}}\xspace}
\def\Dzb     {{\ensuremath{\Dbar{}^0}}\xspace}

\def\Dstarp  {{\ensuremath{\D^{*+}}}\xspace}
\def\Dstarm  {{\ensuremath{\D^{*-}}}\xspace}
\def\Dstarpm {{\ensuremath{\D^{*\pm}}}\xspace}

\def\Bbar    {{\ensuremath{\kern 0.18em\overline{\kern -0.18em \PB}{}}}\xspace}

\def\BorBbar    {\kern 0.18em\optbar{\kern -0.18em B}{}\xspace}

  \def\Y#1S{\ensuremath{\PUpsilon{(#1S)}}\xspace}% no space before {...}!

\def\Lbar        {{\ensuremath{\kern 0.1em\overline{\kern -0.1em\PLambda}}}\xspace}
\def\LorLbar    {\kern 0.18em\optbar{\kern -0.18em \PLambda}{}\xspace}

\newcommand{\decay}[2]{\ensuremath{#1\!\to #2}\xspace}         % {\Pa}{\Pb \Pc}

\def\to                 {\ensuremath{\rightarrow}\xspace}

\def\CP                {{\ensuremath{C\!P}}\xspace}

\newcommand{\dm}{{\ensuremath{\Delta m}}\xspace}

\newcommand{\Delm}{{\mbox{$\Delta m $}}\xspace}
\newcommand{\ACP}{{\ensuremath{{\mathcal{A}}^{\CP}}}\xspace}

\newcommand{\DACP}{{\ensuremath{{\Delta\mathcal{A}}^{\CP}}}\xspace}
\newcommand{\Arawsig}{{\ensuremath{{\mathcal{A}}^{\mathrm{raw}}_{\mathrm{sig}}}}\xspace}
\newcommand{\Arawbg}{{\ensuremath{{\mathcal{A}}^{\mathrm{raw}}_{\mathrm{bkg}}}}\xspace}
\newcommand{\Araw}{{\ensuremath{{\mathcal{A}}^{\mathrm{raw}}}}\xspace}
\newcommand{\Adet}{{\ensuremath{{\mathcal{A}}^{\mathrm{det}}}}\xspace}
\newcommand{\Aprod}{{\ensuremath{{\mathcal{A}}^{\mathrm{prod}}}}\xspace}

\newcommand{\pitag}{\ensuremath{\pi_{\mathrm{tag}}}\xspace}
\newcommand{\pitagp}{\ensuremath{\pi_{\mathrm{tag}}^{+}}\xspace}
\newcommand{\pitagm}{\ensuremath{\pi_{\mathrm{tag}}^{-}}\xspace}

\def\AT#1     {\ensuremath{A_{\mathrm{T}}^{#1}}\xspace}           % 2

\def\C#1      {\ensuremath{\mathcal{C}_{#1}}\xspace}                       % 9
\def\Cp#1     {\ensuremath{\mathcal{C}_{#1}^{'}}\xspace}                    % 7
\def\Ceff#1   {\ensuremath{\mathcal{C}_{#1}^{\mathrm{(eff)}}}\xspace}        % 9  
\def\Cpeff#1  {\ensuremath{\mathcal{C}_{#1}^{'\mathrm{(eff)}}}\xspace}       % 7
\def\Ope#1    {\ensuremath{\mathcal{O}_{#1}}\xspace}                       % 2
\def\Opep#1   {\ensuremath{\mathcal{O}_{#1}^{'}}\xspace}                    % 7

\newcommand{\tev}{\ifthenelse{\boolean{inbibliography}}{\ensuremath{~T\kern -0.05em eV}}{\ensuremath{\mathrm{\,Te\kern -0.1em V}}}\xspace}
\newcommand{\gev}{\ensuremath{\mathrm{\,Ge\kern -0.1em V}}\xspace}
\newcommand{\mev}{\ensuremath{\mathrm{\,Me\kern -0.1em V}}\xspace}
\newcommand{\kev}{\ensuremath{\mathrm{\,ke\kern -0.1em V}}\xspace}
\newcommand{\ev}{\ensuremath{\mathrm{\,e\kern -0.1em V}}\xspace}
\newcommand{\gevc}{\ensuremath{{\mathrm{\,Ge\kern -0.1em V\!/}c}}\xspace}
\newcommand{\mevc}{\ensuremath{{\mathrm{\,Me\kern -0.1em V\!/}c}}\xspace}
\newcommand{\gevcc}{\ensuremath{{\mathrm{\,Ge\kern -0.1em V\!/}c^2}}\xspace}
\newcommand{\gevgevcccc}{\ensuremath{{\mathrm{\,Ge\kern -0.1em V^2\!/}c^4}}\xspace}
\newcommand{\mevcc}{\ensuremath{{\mathrm{\,Me\kern -0.1em V\!/}c^2}}\xspace}

\def\mum  {\ensuremath{{\,\upmu\mathrm{m}}}\xspace}

\def\invfb   {\ensuremath{\mbox{\,fb}^{-1}}\xspace}

\newcommand{\chisq}{\ensuremath{\chi^2}\xspace}

\newcommand{\chisqip}{\ensuremath{\chi^2_{\text{IP}}}\xspace}

\newcommand{\chisqfd}{\ensuremath{\chi^2_{\text{FD}}}\xspace}

\def\gsim{{~\raise.15em\hbox{$>$}\kern-.85em
          \lower.35em\hbox{$\sim$}~}\xspace}
\def\lsim{{~\raise.15em\hbox{$<$}\kern-.85em
          \lower.35em\hbox{$\sim$}~}\xspace}

\def\ptot       {\mbox{$p$}\xspace}
\def\pt         {\mbox{$p_{\mathrm{ T}}$}\xspace}

\def\evtgen     {\mbox{\textsc{EvtGen}}\xspace}

\def\geant      {\mbox{\textsc{Geant4}}\xspace}

\def\photos     {\mbox{\textsc{Photos}}\xspace}

\def\pythia     {\mbox{\textsc{Pythia}}\xspace}

\def\tell1  {TELL1\xspace}
\def\ukl1   {UKL1\xspace}

\usepackage{cite} % Allows for ranges in citations
\usepackage{mciteplus}

\usepackage{longtable} % only for template; not usually to be used in PAPERs

\begin{document}

%%%%%%%%%%%%%%%%%%%%%%%%%
%%%%% Title     %%%%%%%%%
%%%%%%%%%%%%%%%%%%%%%%%%%
\renewcommand{\thefootnote}{\fnsymbol{footnote}}
\setcounter{footnote}{1}

% %%%%%%% CHOOSE TITLE PAGE--------
%\onecolumn
% $Id: title-LHCb-PAPER.tex 111203 2017-08-08 15:28:40Z pkoppenb $
% ===============================================================================
% Purpose: LHCb-PAPER journal paper title page template
% Author: 
% Created on: 2010-09-25
% ===============================================================================

%%%%%%%%%%%%%%%%%%%%%%%%%
%%%%%  TITLE PAGE  %%%%%%
%%%%%%%%%%%%%%%%%%%%%%%%%
\begin{titlepage}
\pagenumbering{roman}

% Header ---------------------------------------------------
\vspace*{-1.5cm}
\centerline{\large EUROPEAN ORGANIZATION FOR NUCLEAR RESEARCH (CERN)}
\vspace*{1.5cm}
\noindent
\begin{tabular*}{\linewidth}{lc@{\extracolsep{\fill}}r@{\extracolsep{0pt}}}
\ifthenelse{\boolean{pdflatex}}% Logo format choice
{\vspace*{-1.5cm}\mbox{\!\!\!\includegraphics[width=.14\textwidth]{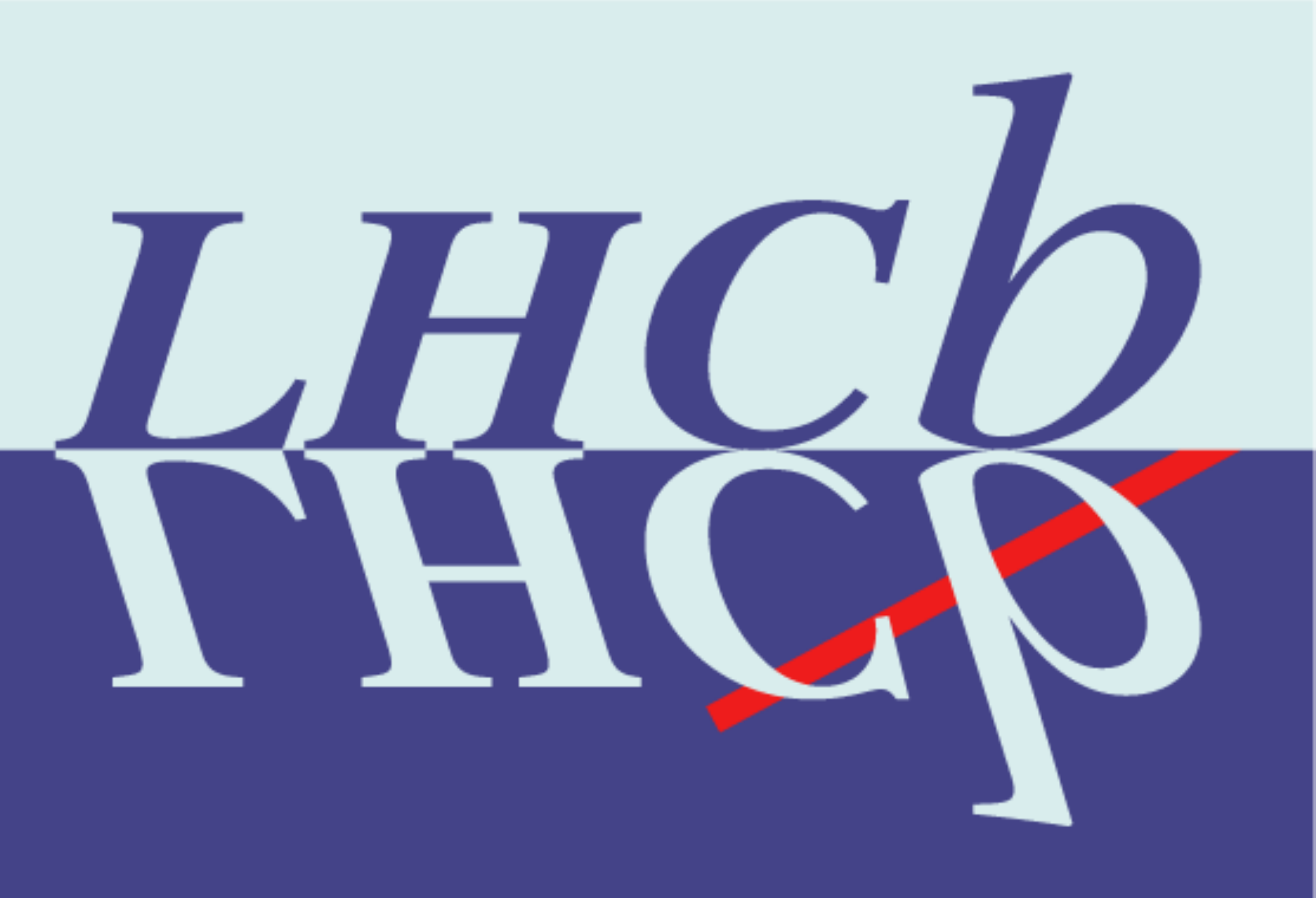}} & &}%
{\vspace*{-1.2cm}\mbox{\!\!\!\includegraphics[width=.12\textwidth]{lhcb-logo.eps}} & &}%
\\
 & & CERN-EP-2018-133 \\  % ID 
 & & LHCb-PAPER-2018-012 \\  % ID 
 & & 20 November 2018 \\ % Date - Can also hardwire e.g.: 23 March 2010
 & & \\
% not in paper \hline
\end{tabular*}

\vspace*{3.0cm}

% Title --------------------------------------------------
{\normalfont\bfseries\boldmath\huge
\begin{center}
% DO NOT EDIT HERE. Instead edit macro in main.tex to keep metadata correct
  \papertitle 
\end{center}
}

\vspace*{2.0cm}

% Authors -------------------------------------------------
\begin{center}
%In the footnote, replace 'paper' by 'Letter' in case of submission to PRL or PLB 
% Edit macro in main.tex to keep metadata correct
\paperauthors\footnote{Authors are listed at the end of this paper.}
\end{center}

\vspace{\fill}

% Abstract -----------------------------------------------
\begin{abstract}
  \noindent
A measurement of the time-integrated \CP\ asymmetry in $D^0\rightarrow
\KS\KS$ decays is reported. The data correspond to an integrated luminosity of about
$2\invfb$ collected in 2015--2016 by the LHCb collaboration in $pp$ collisions at a
centre-of-mass energy of $13\tev$. 
The \Dz candidate is required to originate from a $\Dstarp\to D^0 \pip$ decay, allowing
the determination of the flavour of the \Dz meson using the pion charge.
The $\Dz \to K^{+}K^{-}$ decay, which has a well measured \CP asymmetry, is
used as a calibration channel.
The \CP asymmetry for $\Dz\to\KS\KS$ is measured to be
  \[ \ACP(D^0\rightarrow \KS\KS) = (4.3\pm 3.4\pm 1.0)\%, \]
where the first uncertainty is statistical and the second is systematic. 
This result is combined with the previous LHCb measurement at lower centre-of-mass
energies to obtain
  \[ \ACP(D^0\rightarrow \KS\KS) = (2.3\pm 2.8\pm 0.9)\%. \]

\end{abstract}

\vspace*{1.0cm}

\begin{center}
Published in JHEP 11 (2018) 048
\end{center}

\vspace{\fill}

{\footnotesize 
	% Edit macro in main.tex to keep metadata correct
	\centerline{\copyright~\papercopyright. \href{\paperlicenceurl}{\paperlicence}.}}
\vspace*{2mm}

\end{titlepage}

%%%%%%%%%%%%%%%%%%%%%%%%%%%%%%%%
%%%%%  EOD OF TITLE PAGE  %%%%%%
%%%%%%%%%%%%%%%%%%%%%%%%%%%%%%%%

%  empty page follows the title page ----
\newpage
\setcounter{page}{2}
\mbox{~}
%\newpage
%
%% Author List ----------------------------
%%  You need to get a new author list!
%\input{LHCb_authorlist.tex}
%
%The author list for journal publications is provided by the Membership Committee shortly after 'approval to go to paper' has been given.
%%It will be made available on the page
%%\verb!http://www.physik.uzh.ch/~strauman/forMemCo/LHCb-PAPER-XXXX-XXX/! .
%It will be sent to you by email shortly after a paper number has beens assigned.
%The author list should be included already at first circulation, 
%to allow new members of the collaboration to verify whether they have been included correctly.
%Occasionally a misspelled name is corrected or associated institutions become full members.
%In that case, a new author list will be sent to you.
%In case line numbering doesn't work well after including the authorlist, try moving the \verb!\bigskip! after the last author to a separate line.
%
%
%The authorship for Conference Reports should be ``The LHCb
%  collaboration'', with a footnote giving the name(s) of the contact
%  author(s), but without the full list of collaboration names.

\cleardoublepage

%\twocolumn
% %%%%%%%%%%%%% ---------

\renewcommand{\thefootnote}{\arabic{footnote}}
\setcounter{footnote}{0}

%%%%%%%%%%%%%%%%%%%%%%%%%%%%%%%%
%%%%%  Table of Content   %%%%%%
%%%%%%%%%%%%%%%%%%%%%%%%%%%%%%%%
%%%% Uncomment next 2 lines if desired
%\tableofcontents
%\cleardoublepage

%%%%%%%%%%%%%%%%%%%%%%%%%
%%%%% Main text %%%%%%%%%
%%%%%%%%%%%%%%%%%%%%%%%%%

\pagestyle{plain} % restore page numbers for the main text
\setcounter{page}{1}
\pagenumbering{arabic}

%\linenumbers

% $Id: introduction.tex 87303 2016-02-08 13:44:29Z lafferty $

\section{Introduction}
\label{sec:Introduction}
In the Standard Model, violation of charge-parity (\CP) symmetry originates from
the presence of a single phase in the Cabibbo-Kobayashi-Maskawa (CKM)
matrix~\cite{doi:10.1143}. Experimental results support the CKM mechanism for \CP violation, 
but additional sources  of \CP violation are needed to explain
cosmological observations of the relative abundance of matter and antimatter in the
universe~\cite{Dine:2003ax}. 
In the charm sector, \CP violation has not yet been observed, but measurements of 
\CP asymmetries in Cabibbo-suppressed $\Dz \to h^+ h^{-}$ decays ($h= \pi, K$) have
reached 0.2\% and 0.03\% precision for 
time-integrated~\cite{LHCb-PAPER-2016-035} and indirect \CP asymmetries~\cite{LHCb-PAPER-2016-063}, 
respectively.

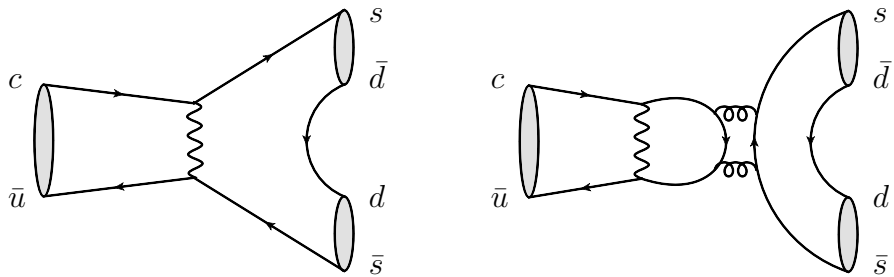
\begin{figure}[h]
 \centering 
 \includegraphics[scale=0.35]{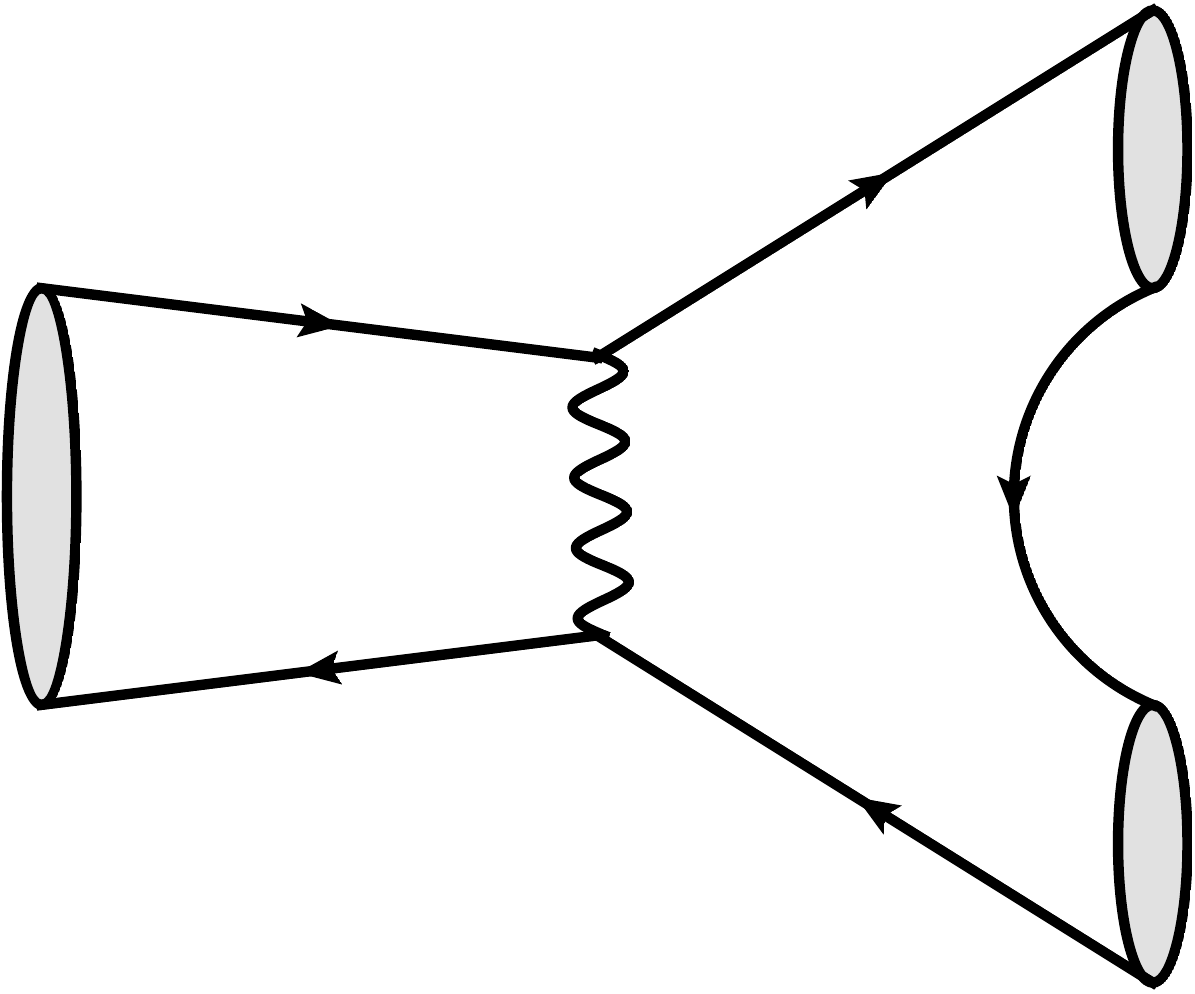}\put(-130,25){$\bar{u}$}\put(-130,70){$c$}\put(5,95){$s$}\put(5,70){$\bar{d}$}\put(5,25){$d$}\put(5,0){$\bar{s}$}
 \hspace{2cm}
 \includegraphics[scale=0.35]{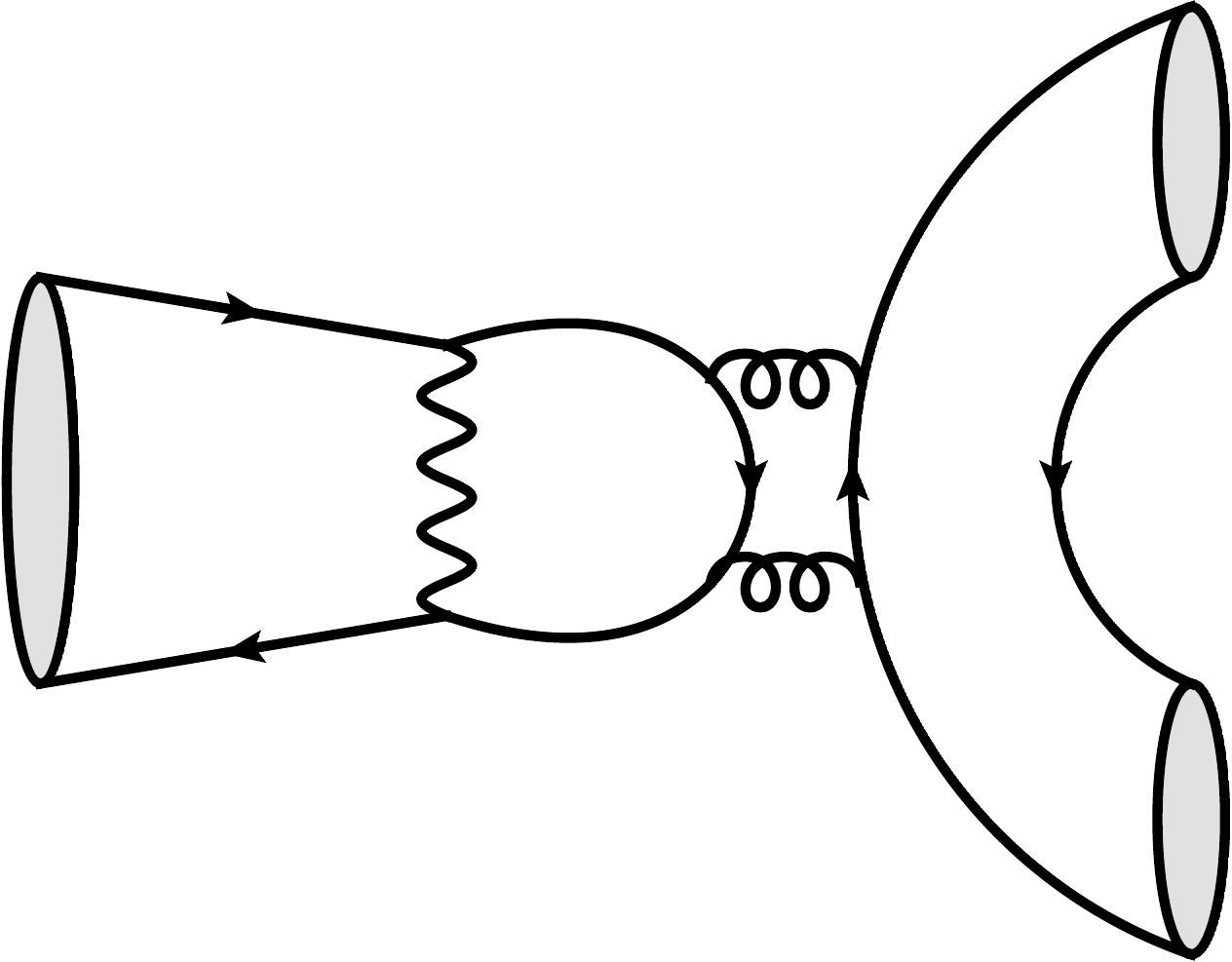}\put(-138,25){$\bar{u}$}\put(-138,70){$c$}\put(5,95){$s$}\put(5,70){$\bar{d}$}\put(5,25){$d$}\put(5,0){$\bar{s}$}
\caption{Exchange (left) and penguin annihilation (right) diagrams contributing to the $\Dz\to\KS\KS$ amplitude. Based on Ref.~\cite{Nierste:2015zra}.} 
\label{fig:diagrams}
\end{figure}

The $\Dz \to \KS \KS$ decay is a promising discovery channel for \CP violation
in charm decays~\cite{Nierste:2015zra}. 
% In fact, because of the
% presence of suppressed penguin contributions and SU(3) breaking contributions only, 
Only loop-suppressed amplitudes and exchange 
diagrams that vanish in the SU(3) flavour limit contribute to this decay. 
These amplitudes can have different strong and weak phases and are of similar size. The time-integrated
\CP asymmetry, $\ACP$, in $\Dz \to \KS \KS$ decays may therefore be enhanced to an observable
level~\cite{Brod:2011re}, and could be as large as 1.1\%~\cite{Nierste:2015zra}.  
Examples of such diagrams are shown in Fig.~\ref{fig:diagrams}. 
The most precise measurement of this asymmetry to date, $\ACP(\KS\KS) = (-0.02\pm1.53\pm0.17)$\%, has been
performed by the \belle\ collaboration~\cite{Dash:2017heu}. Earlier measurements were 
also performed by the \lhcb~\cite{LHCb-PAPER-2015-030} 
and \cleo~\cite{Bonvicini:2000qm} collaborations. 
This article reports a new measurement of \ACP in the decay $\Dz \to \KS \KS$ using \lhcb data collected in 2015 and 2016. 

%\begin{figure}[h]
% \centering 
% \includegraphics[scale=0.35]{./figs/exchange.eps}\put(-130,25){$c$}\put(-130,70){$\bar{u}$}\put(5,95){$\bar{d}$}\put(5,70){$s$}\put(5,25){$\bar{s}$}\put(5,0){$d$}
% \hspace{2cm}
% \includegraphics[scale=0.35]{./figs/peng_ann.eps}\put(-138,25){$c$}\put(-138,70){$\bar{u}$}\put(5,95){$\bar{d}$}\put(5,70){$s$}\put(5,25){$\bar{s}$}\put(5,0){$d$}
% \hspace{2cm}
%\caption{Exchange (left) and penguin annihilation (right) diagrams contributing to the $\Dz\to\KS\KS$ amplitude.} %~\cite{Muller:2015lua}.
%\label{fig:diagrams}
%\end{figure} %\cite{Grossman:2011zk}

The measurement of the \CP\ asymmetry, defined as
\begin{equation}
\ACP(\KS \KS) \equiv \dfrac{\Gamma(\Dz \to \KS \KS)-\Gamma(\Dzb \to \KS \KS)}{\Gamma(\Dz \to \KS \KS)+\Gamma(\Dzb \to \KS \KS)},
\end{equation}
requires knowledge of the flavour of the \Dz meson at production. A sample of
flavour-tagged $\Dz \to \KS \KS$ decays is obtained by selecting \Dstarp
mesons that are produced in the primary interaction (hereafter referred to as prompt), 
with the subsequent decay $\Dstarp \to \Dz \pip$.\footnote{The
inclusion of charge-conjugate processes is implied throughout this document,
unless explicitly specified.} The charge of the pion in this decay identifies the
flavour of the accompanying \Dz meson. The effect of $\Dz -\Dzb$ mixing~\cite{PDG2017} is negligible compared to the 
precision of this analysis and is not considered further. 

The experimentally measured quantity is the raw asymmetry, defined as 
\begin{equation}
\Araw \equiv \dfrac{N_{\Dz}-N_{\Dzb}}{N_{\Dz}+N_{\Dzb}},
\end{equation}
where $N_{\Dz}$ is the measured yield of $\Dstarp \to \Dz \pip$, $\Dz \to \KS
\KS$ decays and $N_{\Dzb}$ is the measured yield of $\Dstarm \to \Dzb
\pim$, $\Dzb \to \KS \KS$ decays.  This observable is related to the \CP
asymmetry by the expression, valid for small asymmetries,
\begin{equation}
\Araw\approx \ACP + \Aprod + \Adet,
\end{equation}
where \Aprod\ is the \Dstarpm\ production asymmetry, defined as 
$\Aprod \equiv \frac{\sigma(\Dstarp)-\sigma(\Dstarm)}{\sigma(\Dstarp)+\sigma(\Dstarm)}$, 
and \Adet\ is the $\pitag^{\pm}$ detection asymmetry, defined as
$\Adet \equiv \frac{\epsilon(\pitagp)-\epsilon(\pitagm)}{\epsilon(\pitagp)+\epsilon(\pitagm)}$.
The symbol $\pitag^\pm$ refers to the pion in the \Dstarpm decay. To a very good approximation, 
knowledge of \Adet\ and \Aprod\ is unnecessary when using a calibration channel with the 
same production and tagging mechanism. The decay channel $\Dz \to K^{+}K^{-}$ is used for 
this purpose.
% To estimate the production
%and detection asymmetries, the calibration channel $\Dz \to K^{+}K^{-}$ is used,
%where the $\Dz$ is still tagged with the $\Dstarp \to \Dz \pip$ decay.   
The production and detection asymmetries cancel when taking the difference of the
raw asymmetries:
\begin{eqnarray}
\Delta \ACP & \equiv  & \Araw(\KS \KS) - \Araw(K^{+}K^{-}) \\
            & =       & \ACP(\KS\KS) - \ACP(K^+K^-) \label{eq:dacp}.
\end{eqnarray}
The quantity $\ACP(K^+K^-)$ has been measured with a precision of 0.2\%~\cite{LHCb-PAPER-2016-035},
thus allowing the determination of $\ACP(\KS\KS)$.  

\section{LHCb detector}
\label{sec:Detector}

The \lhcb detector~\cite{Alves:2008zz,LHCb-DP-2014-002} is a single-arm forward
spectrometer covering the \mbox{pseudorapidity} range $2<\eta <5$,
designed for the study of particles containing \bquark or \cquark
quarks. The detector includes a high-precision tracking system
consisting of a silicon-strip vertex detector surrounding the $pp$
interaction region, a large-area silicon-strip detector (TT) located
upstream of a dipole magnet with a bending power of about
$4{\mathrm{\,Tm}}$, and three stations of silicon-strip detectors and straw
drift tubes placed downstream of the magnet.
The tracking system provides a measurement of momentum, \ptot, of charged particles with
a relative uncertainty that varies from 0.5\% at low momentum to 1.0\% at 200\gevc.
The minimum distance of a track to a primary vertex (PV), the impact parameter (IP), 
is measured with a resolution of $(15+29/\pt)\mum$,
where \pt is the component of the momentum transverse to the beam, in\,\gevc.
Different types of charged hadrons are distinguished using information
from two ring-imaging Cherenkov (RICH) detectors. 
Photons, electrons and hadrons are identified by a calorimeter system consisting of
scintillating-pad and preshower detectors, an electromagnetic
calorimeter and a hadronic calorimeter. Muons are identified by a
system composed of alternating layers of iron and multiwire
proportional chambers.
The magnetic field deflects oppositely-charged particles in opposite
directions and this can lead to detection asymmetries. Periodically
reversing the magnetic field polarity throughout the data taking almost cancels
the effect. The configuration with the magnetic field pointing upwards (downwards), 
MagUp (MagDown), bends positively (negatively) charged particles
in the horizontal plane towards the centre of the LHC ring.

The online event selection is performed by a trigger, %~\cite{LHCb-DP-2012-004}\verb!*!
which consists of a hardware stage, based on information from the calorimeter and muon
systems, followed by a software stage, which applies a full event
reconstruction.
At the hardware trigger stage, events are required to have a muon with high \pt or a
hadron, photon or electron with high transverse-energy deposit in the calorimeters. 

Simulated events are used at various phases of the analysis.
In the simulation, $pp$ collisions are generated using
\pythia~\cite{Sjostrand:2006za,*Sjostrand:2007gs} 
with a specific \lhcb
configuration~\cite{LHCb-PROC-2010-056}.  Decays of hadronic particles
are described by \evtgen~\cite{Lange:2001uf}, in which final-state
radiation is generated using \photos~\cite{Golonka:2005pn}. The
interaction of the generated particles with the detector, and its response,
are implemented using the \geant
toolkit~\cite{Allison:2006ve, *Agostinelli:2002hh} as described in
Ref.~\cite{LHCb-PROC-2011-006}.

\section{Event selection}
\label{sec:selection}
The 2015 and 2016 data samples collected in $pp$ collisions at $13\tev$, which
correspond to about $2\invfb$ of integrated luminosity, are used in this analysis. Candidates are reconstructed in the decay $\Dstarp\to\Dz\pi^+$, followed by $\Dz\to\KS\KS$ and then
\decay{\KS}{\pip\pim}. 
The hardware trigger decision is required to be based either on the 
transverse energy deposited in the hadronic calorimeter by a charged particle 
from the decay of the $\Dz$ meson, 
or on signatures not associated with the $\Dstarp$ decay, such as a high-\pt muon, or a high transverse-energy deposit in the electromagnetic or hadronic calorimeters.
The first stage of the software trigger selects a sample with enhanced heavy-flavour content 
by requiring the presence of a large IP, high-\pt charged particle. 
In the second stage of the software trigger, each selected event 
is required to contain at least one fully-reconstructed candidate 
for the $\Dstarp\to\Dz\pi^+$, $\Dz\to\KS\KS$ decay. 

The decays \decay{\KS}{\pip\pim} are reconstructed in two different categories:
the first involving \KS mesons that decay early enough for the
decay products to be reconstructed in the vertex detector; and the
second containing \KS candidates that decay outside the acceptance of the vertex detector, but within the TT acceptance. These categories are
referred to as \emph{long} and \emph{downstream}, respectively. The
long category has better mass, momentum and decay-vertex resolution than the
downstream category.
In this analysis at least one \KS in each \Dz decay is required to be of the long type.
There are therefore two subsamples used: one where both
\KS candidates are long and the other where one is long and the other
is downstream. These are referred to as the LL and LD subsamples, and are analysed separately, since they exhibit different resolutions.
One or more of the charged decay products from a long \KS meson is required to activate
the first stage of the software trigger. 
The pion candidates used in the \KS\ reconstruction are required to be
high-quality tracks, using the $\chisq$/ndf of the track fit and 
the output ${\cal P}_{\rm fake}$ of a multivariate classifier, trained to identify fake tracks, 
that combines information from the particle identification and tracking systems. 
To ensure that pion candidates do not originate from the PV, they are required to satisfy $\chisqip > 36$. 
The quantity \chisqip\ for a
given particle is defined as the difference in the vertex fit \chisq of the
PV associated to the particle, reconstructed with and without the
particle being considered. For downstream \KS\ candidates, the pions
are required to satisfy $p>3$\gevc\ and $\pt>175$\mevc.

Two oppositely charged pions are used to form \KS\ candidates. The vertex fit
is required to satisfy $\chisq<30$ and the \chisqip\ is required to be greater than 9 (4) for long
(downstream) \KS\ candidates. 
Furthermore,  long (downstream) \KS\ candidates are required to satisfy
$\pt>500\ (750)$\mevc.  

Two reconstructed \KS\ candidates are paired to form \Dz\ candidates, requiring
$\chisq < 10$ for the vertex fit. The sum of the \pt\ of the \KS\ candidates is required to
exceed 1500 (2000)\,\mevc\ for LL (LD) candidates. The angle between the \Dz\ momentum
and the vector connecting the PV to the \Dz\ decay vertex is required to be less than 34.6 mrad. The
measured decay time of the \Dz\ meson is required to be greater than 0.2\,ps.  
Finally, the \Dz\ mass is required to be within 20\,\mevcc of the known value~\cite{PDG2017}. 

A pion candidate ($\pitag^{+}$) is added to a reconstructed \Dz\ meson to form a \Dstarp\
candidate, with a \Dstarp\ vertex fit which is required to have $\chisq<25$. 
The $\pitag^{+}$ candidate is required to have $\pt > 100$\,\mevc, 
and to pass through regions of the detector 
that are known to have a small detector asymmetry~\cite{LHCb-PAPER-2015-030}. 
A small fraction of $\pitag^\pm$ candidates are reconstructed with
the wrong charge assignment, and are removed by 
a selection on track quality. 

\begin{figure}[b]
%\begin{center}
\centering
\includegraphics[width=.75\textwidth]{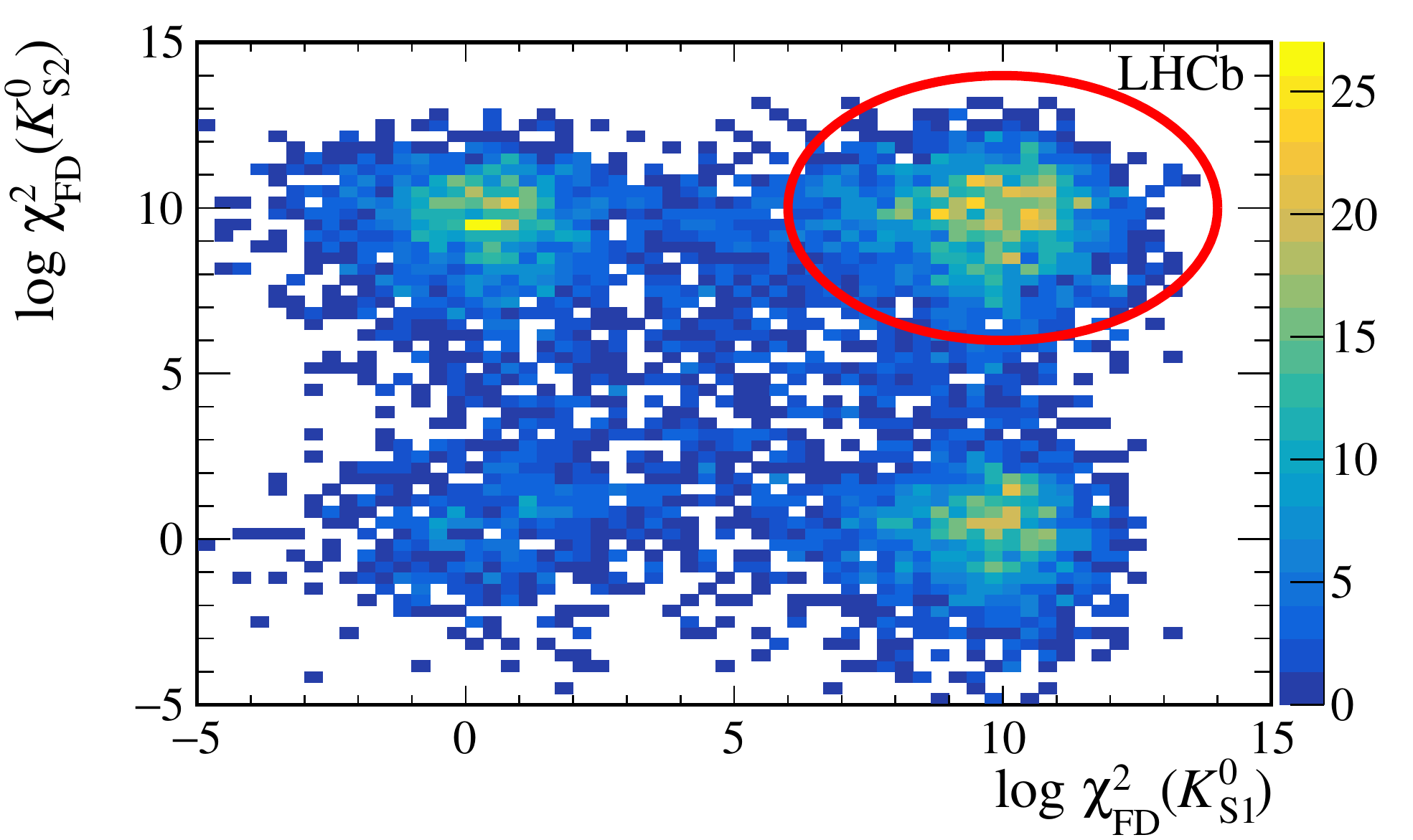}
%\end{center}
\caption{Two-dimensional distribution of the logarithm of the \KS\ flight 
distance significance ($\log \chisqfd$) for
the two \KS\ candidates in the LL subsample of $\Dz\to\KS\KS$ decays. 
The \mbox{$\Dz\to\KS\KS$} signal can be observed in 
the upper right region of the plot. The contour corresponds to Eq.~\ref{eq:fd}.
\label{fig:chisq-fd}
}
\end{figure}

An important source of background is due to the presence of $\Dz \to \KS \pip \pim$
decays, where the \pip \pim pair satisfies the \KS selection. In principle, the contribution of
this channel can be substantial, due to its large branching fraction, but it
is effectively reduced by placing a requirement on the \KS flight distance (FD) and on the mass of the \KS\ candidates. 
The quantity \chisqfd\ is the square of the measured \KS flight distance divided by the square
of its uncertainty. 
Figure~\ref{fig:chisq-fd} shows
a two-dimensional plot of the value of the quantity $\log \chisqfd $ for \KS\ pairs in the LL sample. 
In the figure, four separate regions are visible. The upper right part of the plot, 
where both \KS\ candidates have significant flight distances, is the $\Dz\to\KS\KS$ signal, while the upper left and
lower right regions correspond to $\Dz\to\KS\pip\pim$ decays. The lower left is populated by
$\Dz\to\pip\pim\pip\pim$ decays and combinatorial background.  
A requirement on \chisqfd\ is only necessary for long \KS\ candidates, since downstream \KS\ 
candidates decay far from the PV by construction. For the LL subsample the requirement on the two \KS candidates ($\KSsub{1}$ and $\KSsub{2}$) is
\begin{equation}\label{eq:fd}
[\log \chisqfd (\KSsub{1}) - 10]^2 + [\log \chisqfd (\KSsub{2})-10]^2 < 16, 
\end{equation}
while for the LD sample $\log \chisqfd(\KSsub{L}) > 2.5$ is imposed on the long \KS\ candidate.

The \KS\ mass requirements are 
\begin{equation}
\sqrt{ [m(\KSsub{1})-m_{K^0}]^2 + [m(\KSsub{2})-m_{K^0}]^2 } < 10.5\,\mevcc,
\end{equation}
for LL candidates, with $m_{K^0}=497.6\,\mevcc$~\cite{PDG2017}, and 
\begin{equation}
\sqrt{ \left[\frac{m(\KSsub{L})-m_{K^0}}{10.5\mevcc}\right]^2 + 
       \left[\frac{m(\KSsub{D})-m_{K^0}}{15 \mevcc}\right]^2 } < 1,
\end{equation}
for LD candidates. 
This selection takes into account the difference in resolution between $m(\KSsub{L})$ and $m(\KSsub{D})$. 
The $\log \chisqfd(\KS)$ and $m(\KS)$ regions corresponding to signal and peaking-background candidates are identified using simulations. 
They are further optimised on charge-integrated data by minimising the expected statistical uncertainty on \Araw.

Events in which the \Dstarp meson is not produced in the primary interaction, but
instead is the product of a \bquark-hadron decay, are characterised by a
different production asymmetry and are treated as background. 
These so-called secondary \Dstarp\ candidates tend to have larger values of $\chisqip(\Dz)$ than 
prompt \Dstarp\ candidates and are suppressed by requiring 
$\log \chisqip (\Dz) < 3.0$ (3.5) for the LL (LD) subsample.  
The requirement
$\log \chisqip (\pitag^{+}) < 2.5$ is imposed on both subsamples. Simulated events are used to
estimate the residual secondary fraction in the LL and LD subsamples to be 9\% and 13\%,  
respectively.  

A multivariate classifier, based on the k-nearest neighbours (kNN)
algorithm~\cite{Narsky:2014fya}, is used to further suppress combinatorial
background.  The kNN algorithm classifies events according to the fraction of signal
events among its $k$ nearest neighbours (taken from the training sample of signal and
background events), where the distance is calculated in the $n$-dimensional
space of the input variables and $k$ is a positive integer.  The training
sample uses simulated events for the signal and data events from the \Dz\ mass
sidebands for the background. A wide range of input variables based on track
and vertex quality, the transverse momenta of \KS\ and \Dz\
candidates, helicity angles of the \KS\ and \Dz\ decays and particle
identification information on the pions in the \Dz\ decays was initially considered.
Variables depending on the $\pitag^\pm$ track are not included in the classifier to
avoid introducing possible bias on the asymmetry measurement.  The actual
variables used, the value of $k$, and the selection on the classifier output are optimised separately for the LL
and LD subsamples, using the expected statistical uncertainty on the raw asymmetry as a
figure of merit. 

For the $\Dz\to K^+K^-$ control channel, an attempt is made to keep the
selection similar to the $\Dz\to\KS\KS$ channel, although some
selections made at the software trigger level are different for the two
channels.  
Charged tracks positively identified as kaons in the RICH detectors are selected to 
reconstruct \Dz\ candidates. The kaons are required to satisfy $\chisqip > 4$. For the \Dz\ 
candidates, at least one of the kaons is required to have $\pt > 1$\,\gevc. The sum of the 
kaon momenta is required to exceed 5\,\gevc and the \Dz\ \pt\ is required to be at least 1\,\gevc. Furthermore,
the angle between the \Dz\ momentum vector and the vector connecting the primary and decay vertices  
is required to be less than 17.3 mrad. The following selections are the same as for the 
$\Dz\to\KS\KS$ channel: $\pitag^\pm$ fiducial cuts, fake-track probability and \chisqip\ selection;
and requirements on \Dz\ \chisqip\ and invariant mass.

\section{Asymmetry measurement}
\label{sec:measurement}

The raw asymmetry for $\Dz\to\KS\KS$ is determined by separating the selected
candidates into subsets tagged by positively and negatively charged pions.  A
simultaneous unbinned maximum likelihood fit to their \dm\ distributions is performed, where
\dm\ is the difference of the reconstructed invariant mass of the \Dstarp\ and
the \Dz\ candidates. The calculation of \dm\ is made after the full decay chain
has been reconstructed using a mass constraint on the \KS\ candidates
and constraining the \Dstarp candidate to originate from the PV. 

The signal shape is modelled using the Johnson $S_\mathrm{U}$
distribution~\cite{Johnson:1949zj}, 
which consists of a core Gaussian-like
shape but allows for an asymmetric tail
\begin{equation}\label{eq:signal}
 S(x; \mu,\sigma,\delta,\gamma) \propto  \left[1+\left(\frac{x-\mu}{\sigma}\right)^2\right]^{-\frac{1}{2}} \times \exp\left\{-\frac{1}{2}\left[\gamma + \delta\sinh^{-1}\left(\frac{x-\mu}{\sigma}\right)\right]^2\right\}. 
\end{equation}
The background shape is described
with an exponential function multiplied by a threshold factor and is zero below a fixed endpoint, which is set to the pion mass $m_{\pi}$
\begin{equation}
 B(x; m_{\pi},\chi) \propto \sqrt{x-m_{\pi}} \times \exp{\left(\chi\frac{x}{m_{\pi}}\right)}.
\end{equation}
The likelihood function
is parametrised in terms of \ACP and the expected total number of events $N_\mathrm{exp}=n_\mathrm{sig} + n_\mathrm{bkg}$
\begin{equation}\label{eq:full_likelihood}
\mathcal{L}=\dfrac{e^{-N_\mathrm{exp}}}{N_\mathrm{obs}!}\prod_{i}\left[ n_\mathrm{sig} \dfrac{1+q_i\Arawsig}{2} S(\Delm) + n_\mathrm{bkg} \dfrac{1+q_i\Arawbg}{2} B_{q_i}(\Delm)\right],
\end{equation}
where $n_\mathrm{sig}$ and $n_\mathrm{bkg}$ are the signal and background yields, respectively, and the parameter $q_i = \pm 1$ 
is the charge of the $D^{\ast\pm}$ candidate and $N_{\mathrm{obs}}$ is the total number of candidates.
The signal raw asymmetry \Arawsig is a free parameter in the fit. 
The free parameter \Arawbg\ allows for a possible asymmetry in the combinatorial background.
The four parameters in Eq.~\ref{eq:signal}
defining the signal probability distribution function (PDF) 
are common to the \Dstarp\  and \Dstarm\ samples,
while the parameter describing the background shape is allowed to differ
between the two subsamples.  For the LL sample, there are ten free
parameters. To achieve
convergence of the fit in the smaller LD sample, it is necessary to fix the two parameters that
describe the asymmetric tail in the signal PDF to the values obtained from the
charge-integrated LL subsample. Based on studies of simulated events, the tail
parameters of the LL and LD subsamples are expected to be compatible.  Separate
fits are performed for the two magnet polarities.

\begin{figure}[!htb]
%\begin{center}
\centering
\includegraphics[width=0.48\textwidth]{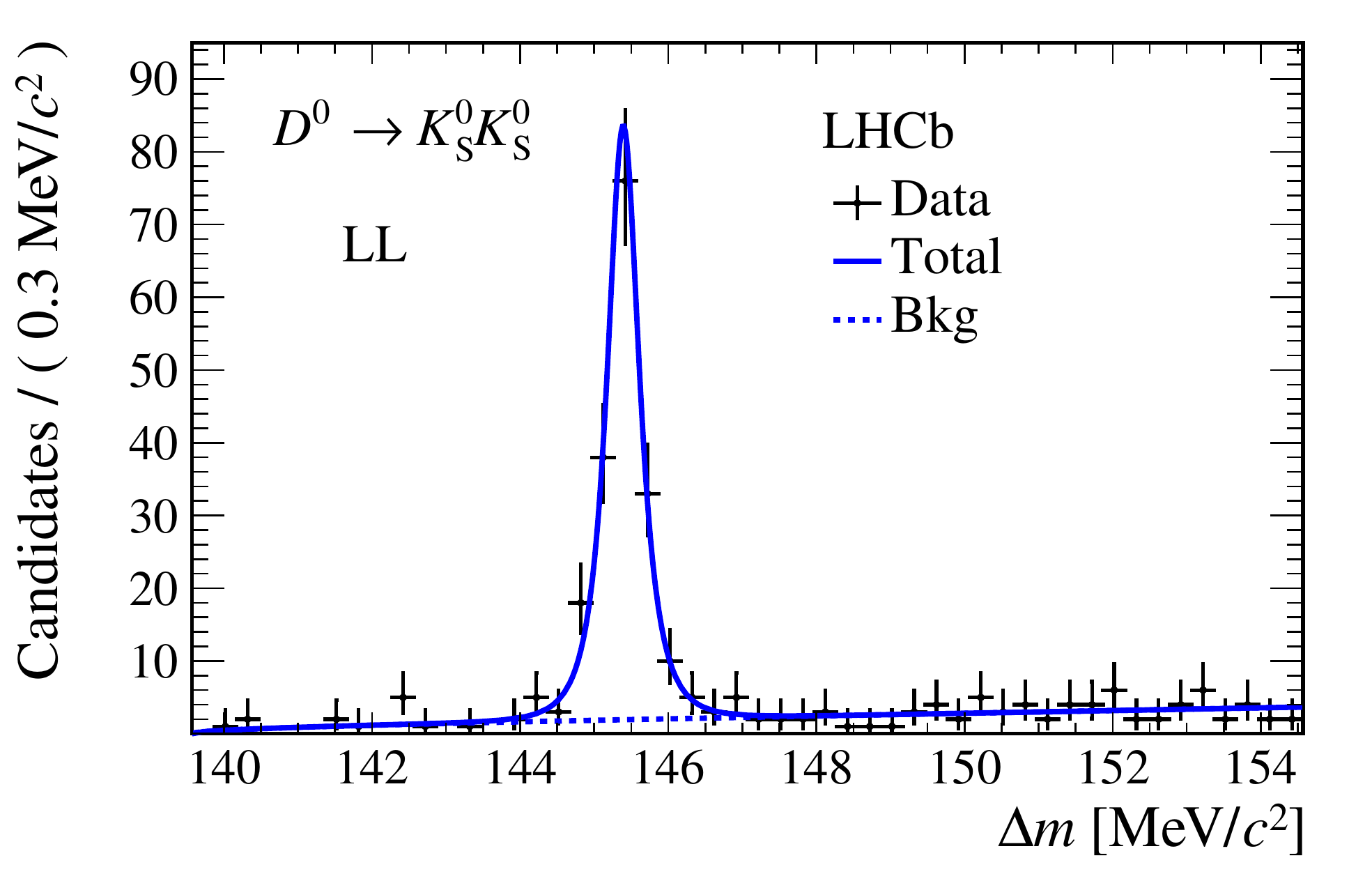}\put(-42,100){(a)}
\includegraphics[width=0.48\textwidth]{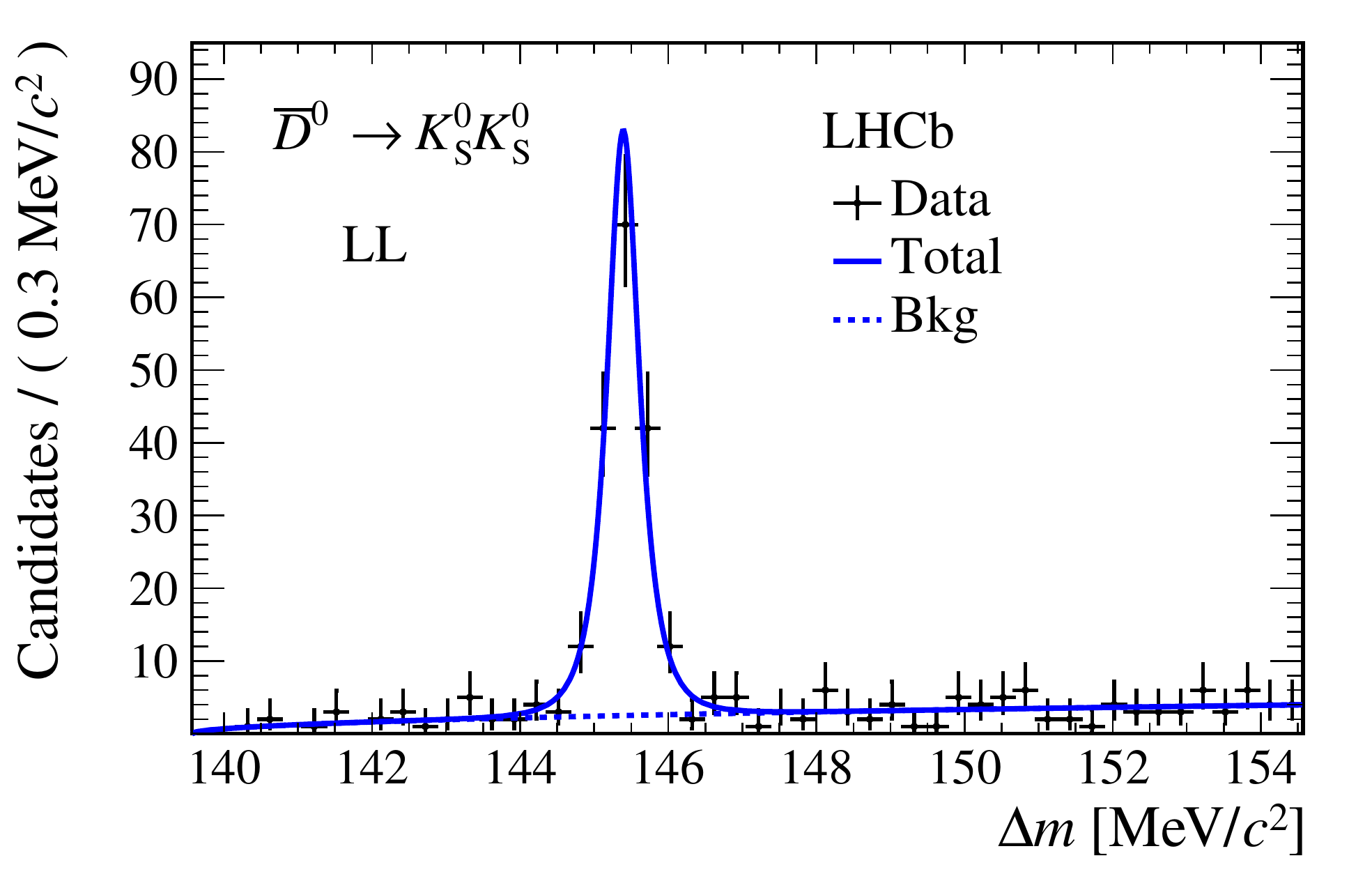}\put(-42,100){(b)}\\
\includegraphics[width=0.48\textwidth]{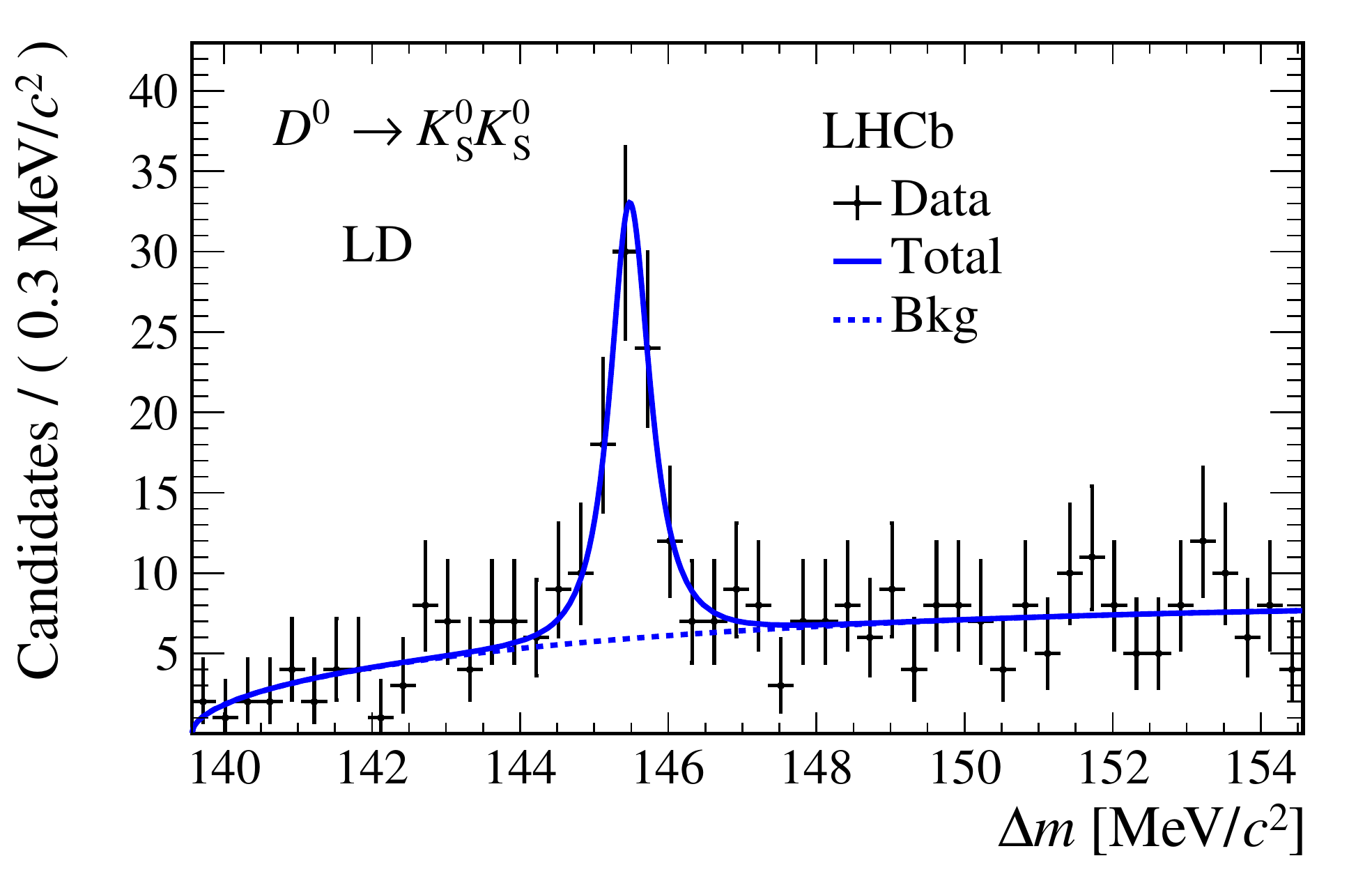}\put(-42,100){(c)}
\includegraphics[width=0.48\textwidth]{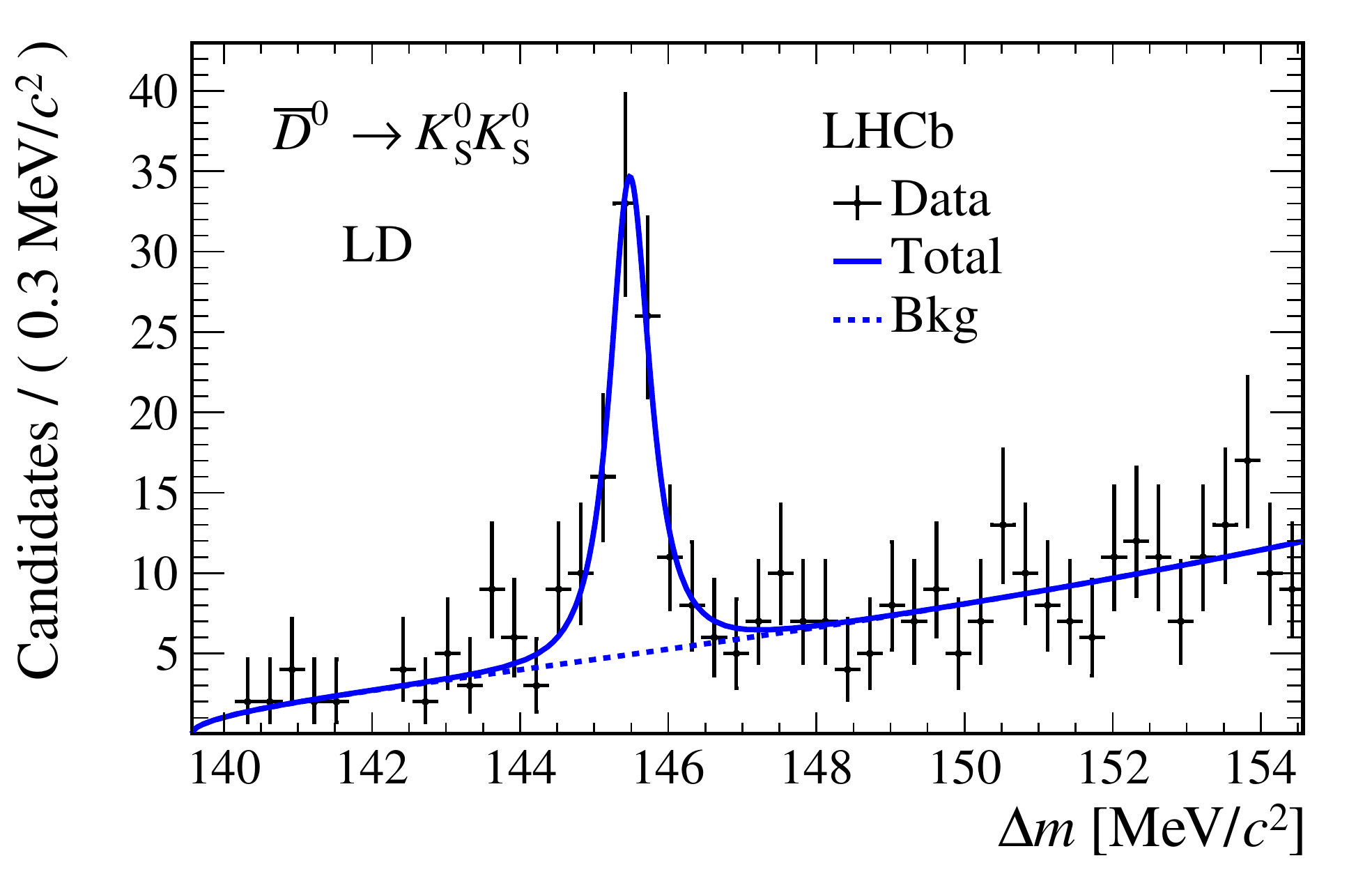}\put(-42,100){(d)}
\caption{Results of fits to \dm\ distributions of $\Dz\to\KS\KS$ candidates for MagUp magnet polarity. 
The fit to (a) $D^{\ast +} \to \Dz \pip$ and (b) $D^{\ast -} \to \Dzb \pim$   candidates for the LL sample and the fit to (c) $D^{\ast +} \to \Dz \pip$ and (d) $D^{\ast -} \to \Dzb \pim$ candidates for the LD sample are shown. The black crosses represent the data points, the solid blue curve is the total fit function, and the 
dashed blue curve is the background component of the fit.
\label{fig:ksks-fit}
}
%\end{center}
\end{figure}

Table~\ref{tab:araw} shows the results of the simultaneous fits to the
$\Dz\to\KS\KS$ candidates. The results on each subset of the data are compatible
with each other.
The fit is shown in Fig.~\ref{fig:ksks-fit} for the samples collected 
with the MagUp magnetic field configuration. 
\begin{table}
\caption{Fit results on the $\Dz\to\KS\KS$ LL and LD samples for each magnet polarity, where $N_{\mathrm{obs}}$ represents the number of
         candidates fitted. The purity is determined in the range \mbox{$ 144.5 < \Delm < 146.5\, \mevcc$}. 
	  For each sample, a $\chisq$ test statistic for the fitted model and 
	  binned data for positively and negatively charged candidates is constructed.
	  The quantity $\mathcal{P}_{\rm fit}$ is the probability of observing a $\chisq$ value greater than 
	  that observed in the fit to real data, determined using simulated pseudoexperiments sampled from the fitted model.  
         } 
\label{tab:araw}
\centering
\resizebox{\columnwidth}{!}{%
\begin{tabular}{lcccccc}
\hline\hline
% Note: \Tspc and \Bspc (defined in main.tex) add a little extra space (top and bottom) to accommodate sub- and superscripts
\Tspc \Bspc & \Arawsig     & $n_\mathrm{sig}$ & \Arawbg     & Purity   & $\mathcal{P}_{\rm fit} (\%)$  & $N_\mathrm{obs}$  \\
\hline
LL MagUp   & $\phantom{-}0.008 \pm 0.057  $ & $ 346 \pm 21 $ & $-0.097 \pm 0.069 $ & 0.92 & 48 & 589 \\
LL MagDown & $\phantom{-}0.103 \pm 0.052  $ & $ 413 \pm 24 $ & $-0.098 \pm 0.068 $ & 0.92 & 43 & 675 \\
LD MagUp   & $-0.046 \pm 0.102 $            & $ 156 \pm 18 $ & $-0.021 \pm 0.044 $ & 0.67 & 93 & 758 \\
LD MagDown & $-0.078 \pm 0.107 $            & $ 152 \pm 19 $ & $-0.040 \pm 0.038 $ & 0.60 & 14 & 950 \\
\hline\hline
\end{tabular}%
}
\end{table}
 
\begin{figure}
%\begin{center}
\centering
\includegraphics[scale=0.38]{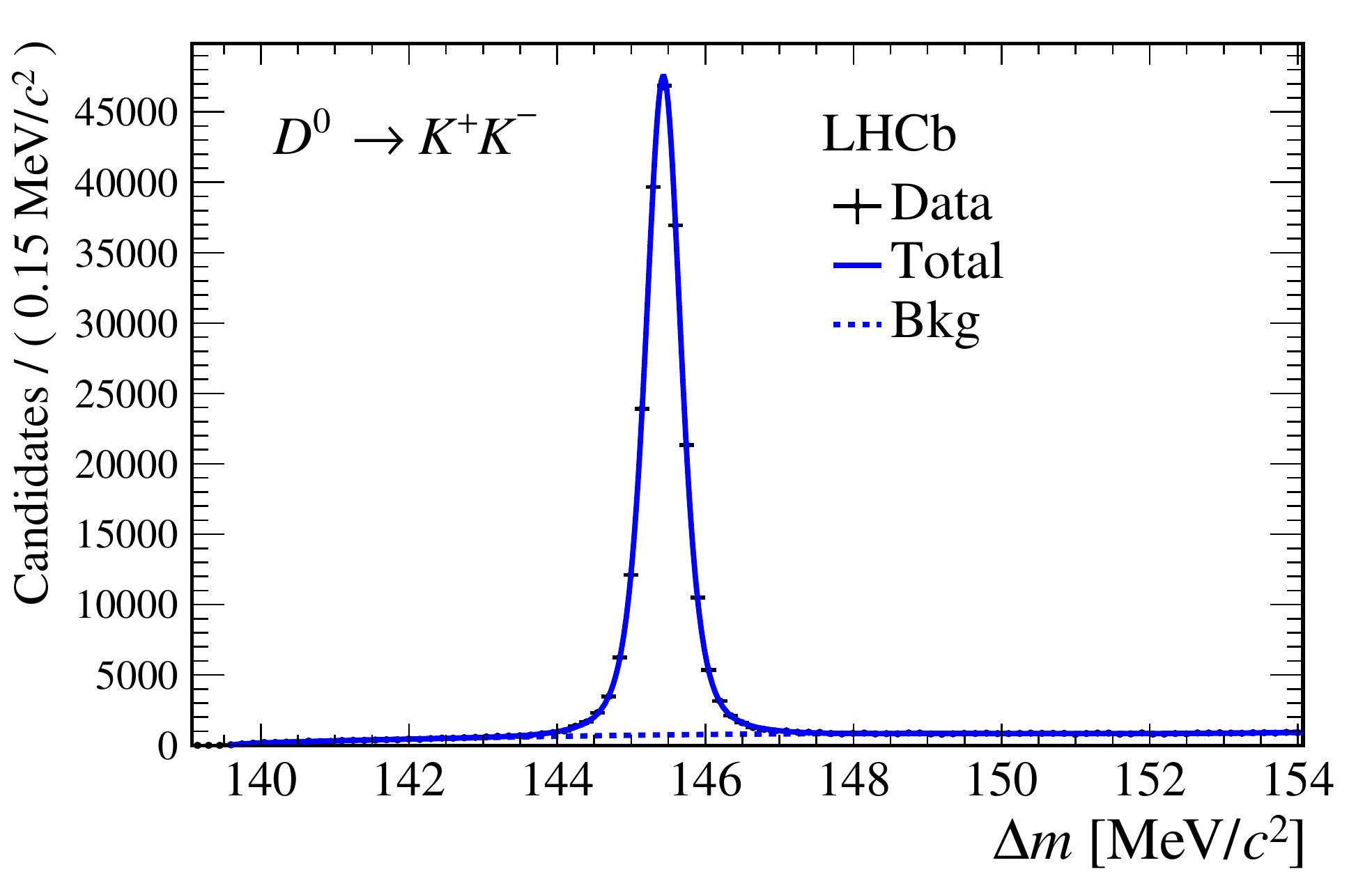}\put(-42,100){(a)}
\includegraphics[scale=0.38]{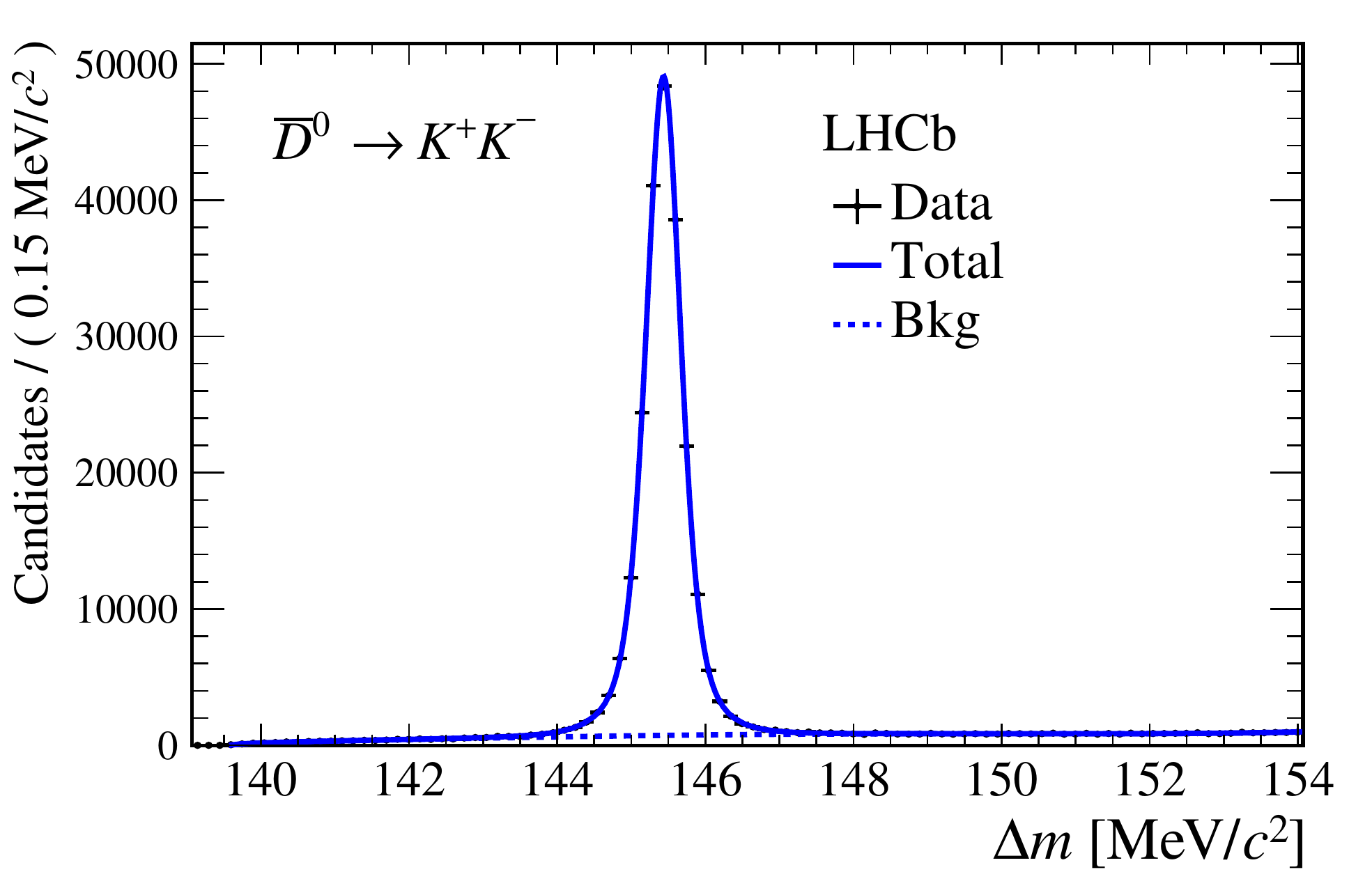}\put(-42,100){(b)}
\caption{Results of fits to \dm\ distributions of $\Dz\to K^+K^-$ candidates for the MagUp magnet polarity. 
The fits to (a) $D^{\ast +} \to \Dz \pip$ candidates and (b)
$D^{\ast -} \to \Dzb \pim$ candidates are shown. The black points represent the data, the dashed blue and solid blue curves
represent the background component and the total fit function, respectively. 
\label{fig:kk-fit}
}
%\end{center}
\end{figure}

For the $\Dz\to K^+K^-$ channel, binned \chisq\ fits are performed to the \dm\ distributions of the positively
and negatively tagged \Dz\ decays. 
The sample consists of $8.25\times 10^5$ selected candidates for the MagDown magnet polarity 
and $5.61\times 10^5$ candidates for the MagUp magnet polarity.
The signal is modelled with a Johnson $S_\mathrm{U}$ distribution plus a Gaussian 
distribution, while the background shape is described by a fourth-degree polynomial multiplied by a 
$\sqrt{\Delm-m_{\pi}}$ threshold factor. 
There are 12 free parameters, and 150 bins, in each \dm\ fit.  
The $\chisq$ probabilities associated to the fits are 
28\% (20\%) for the negatively (positively) tagged \Dz\ decays, and 
23\%  (3\%) for the negatively (positively) tagged \Dz\ decays, in the 
MagUp and MagDown magnet polarities, respectively. 
Figure~\ref{fig:kk-fit} shows the results for
the MagUp magnet polarity fit. The results obtained for the two magnet polarities are 
\begin{eqnarray}\label{eq:ArawKK}
\Araw(K^+K^-)_\mathrm{MagUp\phantom{wn}}   & = &          - 0.0188 \pm 0.0020 , \\
\Araw(K^+K^-)_\mathrm{MagDown}             & = & \phantom{+}0.0030 \pm 0.0017          ,  \nonumber
\end{eqnarray}
where the uncertainties are statistical.
The difference in the MagUp and MagDown values of $\Araw(K^+K^-)$ is an indication of a significant $\pitag^\pm$ detection
asymmetry, which depends on the magnetic field orientation.

\section{Systematic uncertainties}
\label{sec:systematics}

The main source of systematic uncertainty arises from the determination of
\Araw\ on the $\Dz\to\KS\KS$ sample.  Possible bias in the fitting procedure
is evaluated using simulated pseudoexperiments.  In particular, the uncertainty related to 
the choice of the signal model is evaluated by  using the
nominal model to fit samples generated with two alternative models for the signal PDF:
either a sum of two Gaussians with a common mean (for the LL sample) or a single 
Gaussian (for the LD sample). The background PDF is varied by 
modifying its behaviour at threshold. 
Systematic uncertainties of $5\times 10^{-3}$ and 0.01 for the LL and LD samples,
respectively, are assigned based on this study.
As a cross-check, the background shapes are constrained to be the same for the \Dstarp\ and
\Dstarm\ samples, and the resulting asymmetry is compatible with the nominal.
For the $\Dz\to K^+K^-$ fit, an alternative
procedure is used to evaluate the systematic uncertainty associated with the signal PDF. In this case, the signal region ($\pm 2.5$\,\mevcc\ around the signal mean)
is excluded and only the background shape is fit. The yield is then
determined by estimating the background in the signal region by interpolating
the fitted background function.  Additionally, alternative background
shapes are tried, varying the degree of the polynomial.
Based on these studies a systematic uncertainty of $2\times 10^{-3}$  
is assigned to $\Araw(K^+K^-)$.

The contribution of the residual background of $\Dz\to\KS\pi^+\pi^-$ decays to the fitted LL and LD signal yields
is estimated to be $(3.5 \pm 0.7)\%$ and $(5.5 \pm 4.6)\%$, respectively. These values are combined
with the $\KS\pip\pim$ background asymmetry, 
determined from background-dominated regions of the \chisqfd\ distributions, to estimate contributions to
the systematic uncertainty of $4\times 10^{-3}$ and $5\times 10^{-3}$, for the LL and LD samples. 
Another contribution comes from
the residual fraction of secondary decays, 
which leads to a systematic
uncertainty for this source of $2\times 10^{-3}$ and $3\times 10^{-3}$ for the LL and LD samples. In this case
an upper limit of 0.02 for the maximum difference in the
production asymmetries of \Dstarpm\ mesons and
$b$-hadrons is assumed~\cite{LHCb:2012fb,LHCb-PAPER-2014-042,Aaij:2017mso}.  

Potential trigger biases are studied using tagged $\Dz \to \Kp\Km$ decays, 
by comparing the raw asymmetries obtained in the subsample in which the trigger 
decision is based on the charged particles from the decay of the $\Dz$ meson, 
and in the subsample in which the trigger decision is not 
associated with the $\Dstarp$ decay. 
The sum in quadrature of the difference (albeit not statistically significant) 
and of its statistical uncertainty is assigned as a systematic uncertainty, 
which accounts for residual trigger-induced biases in the 
difference of measured asymmetries for signal and control channels.
This uncertainty amounts to $5\times 10^{-3}$ for both the LL and LD samples.
The small probability of assigning the wrong charge to the $\pitag^{\pm}$
candidate results in a systematic uncertainty of $2\times 10^{-3}$ for both the LL and LD
samples. This is obtained by varying the selection on the ${\cal P}_{\rm fake}$ value of  $\pitag^{\pm}$ candidates.
This uncertainty cancels for $\Delta\ACP$.  
For each neutral kaon in the final state, asymmetries arising from regeneration and from mixing 
and \CP\ violation in the $\Kz-\Kzb$ system are suppressed at the ${\cal O}(10^{-3})$ level~\cite{Enz}. 
Since they are expected to affect $\Dz\to\KS\KS$ and $\Dzb\to\KS\KS$ decays by the same amount,  
they cancel in \Araw\ and therefore do not contribute to the systematic uncertainty. 

The cancellation of the production and detection asymmetries in the computation of \DACP\ may not be perfect
due to differences in the kinematics of the $\Dz\to\KS\KS$ candidates and the $\Dz\to K^+K^-$ candidates. 
The offline selection of the two channels aims to keep the kinematics as similar as possible, but the different
trigger selections on the final states
can introduce differences. The associated systematic uncertainty is evaluated 
by considering four kinematic variables: the transverse momentum and the pseudorapidity of the $\Dstarp$ candidate
and the $\pitag^+$ candidate, respectively. For each variable a one-dimensional weighting is performed on the $\Dz\to K^+K^-$ events
such that they have the same distribution as the $\Dz\to\KS\KS$ sample. Then $\Araw(K^+K^-)$ is determined from the weighted
sample. This is repeated for each of the four kinematic variables. The largest change in $\Araw(K^+K^-)$ is taken
as the systematic uncertainty and this is found to be $2\times 10^{-3}$ for both the LL and LD samples.
The systematic uncertainties are summarised in Table~\ref{tab:syst}. 
\begin{table}[tb]
%\begin{center}
\centering
\caption{Systematic uncertainties on the quantities \Araw\ and  $\Delta\ACP$. 
The total systematic uncertainties in the last row are obtained by summing 
the corresponding contributions in each column in quadrature. 
Uncertainties are expressed in units of $10^{-3}$. 
}\label{tab:syst}       
\begin{tabular}{lcccc}
\hline\hline
Source                 & $\Araw({\rm LL})$ & $\Araw({\rm LD})$ & $\Delta\ACP({\rm LL})$ & $\Delta\ACP({\rm LD})$  \\  \hline
Fit procedure          &  5  & 10  & 5  &  10  \\ 
$\KS\pip\pim$ background &  4  &  5  & 4  &  5  \\ 
Secondaries            &  2  &  3  & 2  &  3  \\
Wrong $\pitag^{\pm}$ charge    &  2  &  2  & -- & --  \\
Trigger selection      &  5  &  5  & 5  & 5   \\ 
$K^+K^-$ fit procedure & --  & --  &  2 &  2  \\
%Res. det. asymm.       & --  & --  &  2 &  2  \\   
Residual detection     & \multirow{2}{*}{--}  & \multirow{2}{*}{--}  & \multirow{2}{*}{2} & \multirow{2}{*}{2}  \\   
\hspace{1.5ex} asymmetry              &                      &                      &                    &  \\
\hline 
Total                  & 9   & 13  &  9 & 13  \\ 
\hline\hline
\end{tabular}
%\end{center}
\end{table}

\section{Results}
\label{sec:results}
\begin{figure}
	%\begin{center}
	\centering
	\includegraphics[scale=0.38]{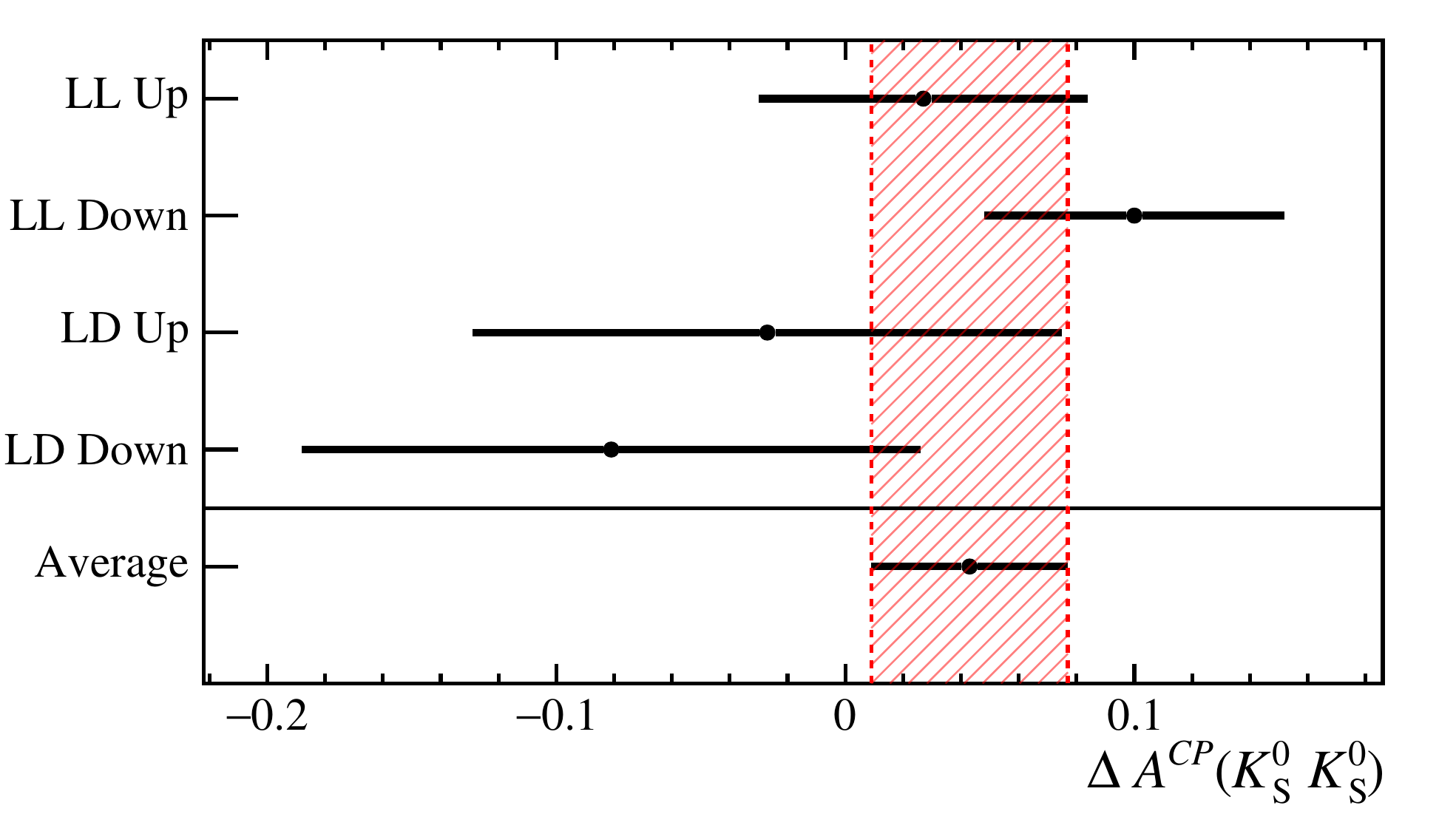}
	\caption{Values of \DACP obtained for both magnet polarities on the LL and LD samples, along with the average of these measurements. Only statistical uncertainties are shown.
		\label{fig:dacp-final}
	}
	%\end{center}
\end{figure}

The procedure described in Sect.~\ref{sec:Introduction} is used to combine the results for the raw 
asymmetries to obtain $\ACP(\KS\KS)$ for each of the LL and LD subsamples. 
For each of the subsamples, the difference \DACP\ is calculated separately for 
the different magnet polarities using the fitted values
of \Araw\ (Table~\ref{tab:araw} and Eq.~\ref{eq:ArawKK}). The values of \DACP\ corresponding to 
the two magnet polarities, which are found to be in good agreement (Fig.~\ref{fig:dacp-final}), are averaged by weighting
with their statistical uncertainties.
The systematic uncertainties are taken from Table~\ref{tab:syst}.    
Using the LHCb measurement of 
$\ACP(K^+K^-) = (0.04\pm0.12\pm0.10)$\%~\cite{LHCb-PAPER-2016-035} results in
\begin{eqnarray*}
\ACP(\mathrm{LL}) & = & \phantom{-}0.067 \pm 0.038 \pm 0.009, \\
\ACP(\mathrm{LD}) & = & -0.053 \pm 0.074 \pm 0.013,
\end{eqnarray*}
where the first uncertainty is statistical and the second is systematic. 
These results are combined by performing an average weighted by the total uncertainties
and assuming that the systematic uncertainties are fully correlated. 
The
final result is
\begin{equation*}
\ACP(\KS\KS) = 0.043 \pm 0.034 \pm 0.010.
\end{equation*}
This measurement is systematically independent of the LHCb Run 1 
measurement, $\ACP(\KS\KS) = -0.029 \pm 0.052 \pm 0.022$~\cite{LHCb-PAPER-2015-030}, 
and is compatible with it. 
An average, weighted by the total uncertainties, of the two measurements is performed to obtain
\begin{equation*}
\ACP(\KS\KS) = 0.023 \pm 0.028 \pm 0.009.
\end{equation*}
These results are compatible with the expectations of the Standard
Model~\cite{Nierste:2015zra} and with previous
measurements~\cite{Bonvicini:2000qm,Dash:2017heu}.

\section*{Acknowledgements}
%
% These Acknowledgements valid from 20-Mar-2018
%
\noindent We express our gratitude to our colleagues in the CERN
accelerator departments for the excellent performance of the LHC. We
thank the technical and administrative staff at the LHCb
institutes. We acknowledge support from CERN and from the national
agencies: CAPES, CNPq, FAPERJ and FINEP (Brazil); MOST and NSFC
(China); CNRS/IN2P3 (France); BMBF, DFG and MPG (Germany); INFN
(Italy); NWO (Netherlands); MNiSW and NCN (Poland); MEN/IFA
(Romania); MinES and FASO (Russia); MinECo (Spain); SNSF and SER
(Switzerland); NASU (Ukraine); STFC (United Kingdom); NSF (USA).  We
acknowledge the computing resources that are provided by CERN, IN2P3
(France), KIT and DESY (Germany), INFN (Italy), SURF (Netherlands),
PIC (Spain), GridPP (United Kingdom), RRCKI and Yandex
LLC (Russia), CSCS (Switzerland), IFIN-HH (Romania), CBPF (Brazil),
PL-GRID (Poland) and OSC (USA). We are indebted to the communities
behind the multiple open-source software packages on which we depend.
Individual groups or members have received support from AvH Foundation
(Germany), EPLANET, Marie Sk\l{}odowska-Curie Actions and ERC
(European Union), ANR, Labex P2IO and OCEVU, and R\'{e}gion
Auvergne-Rh\^{o}ne-Alpes (France), Key Research Program of Frontier
Sciences of CAS, CAS PIFI, and the Thousand Talents Program (China),
RFBR, RSF and Yandex LLC (Russia), GVA, XuntaGal and GENCAT (Spain),
Herchel Smith Fund, the Royal Society, the English-Speaking Union and
the Leverhulme Trust (United Kingdom).

\addcontentsline{toc}{section}{References}
\setboolean{inbibliography}{true}
\bibliographystyle{LHCb}
\bibliography{main,LHCb-PAPER,LHCb-CONF,LHCb-DP,LHCb-TDR,DKSKS}

\ifx\mcitethebibliography\mciteundefinedmacro
\PackageError{LHCb.bst}{mciteplus.sty has not been loaded}
{This bibstyle requires the use of the mciteplus package.}\fi
\providecommand{\href}[2]{#2}
\begin{mcitethebibliography}{10}
\mciteSetBstSublistMode{n}
\mciteSetBstMaxWidthForm{subitem}{\alph{mcitesubitemcount})}
\mciteSetBstSublistLabelBeginEnd{\mcitemaxwidthsubitemform\space}
{\relax}{\relax}

\bibitem{doi:10.1143}
M.~Kobayashi and T.~Maskawa,
  \ifthenelse{\boolean{articletitles}}{\emph{{CP}-violation in the
  renormalizable theory of weak interaction},
  }{}\href{http://dx.doi.org/10.1143/PTP.49.652}{Prog.\ Theor.\ Phys.\
  \textbf{49} (1973) 652}\relax
\mciteBstWouldAddEndPuncttrue
\mciteSetBstMidEndSepPunct{\mcitedefaultmidpunct}
{\mcitedefaultendpunct}{\mcitedefaultseppunct}\relax
\EndOfBibitem
\bibitem{Dine:2003ax}
M.~Dine and A.~Kusenko, \ifthenelse{\boolean{articletitles}}{\emph{{The origin
  of the matter - antimatter asymmetry}},
  }{}\href{http://dx.doi.org/10.1103/RevModPhys.76.1}{Rev.\ Mod.\ Phys.\
  \textbf{76} (2003) 1},
  \href{http://arxiv.org/abs/hep-ph/0303065}{{\normalfont\ttfamily
  arXiv:hep-ph/0303065}}\relax
\mciteBstWouldAddEndPuncttrue
\mciteSetBstMidEndSepPunct{\mcitedefaultmidpunct}
{\mcitedefaultendpunct}{\mcitedefaultseppunct}\relax
\EndOfBibitem
\bibitem{LHCb-PAPER-2016-035}
LHCb collaboration, R.~Aaij {\em et~al.},
  \ifthenelse{\boolean{articletitles}}{\emph{{Measurement of \CP\ asymmetry in
  $\Dz\to\Kp\Km$ decays}},
  }{}\href{http://dx.doi.org/10.1016/j.physletb.2017.01.061}{Phys.\ Lett.\
  \textbf{B767} (2017) 177},
  \href{http://arxiv.org/abs/1610.09476}{{\normalfont\ttfamily
  arXiv:1610.09476}}\relax
\mciteBstWouldAddEndPuncttrue
\mciteSetBstMidEndSepPunct{\mcitedefaultmidpunct}
{\mcitedefaultendpunct}{\mcitedefaultseppunct}\relax
\EndOfBibitem
\bibitem{LHCb-PAPER-2016-063}
LHCb collaboration, R.~Aaij {\em et~al.},
  \ifthenelse{\boolean{articletitles}}{\emph{{Measurement of the \CP\ violation
  parameter $A_\Gamma$ in $\Dz\to\Kp\Km$ and $\Dz\to\pip\pim$ decays}},
  }{}\href{http://dx.doi.org/10.1103/PhysRevLett.118.261803}{Phys.\ Rev.\
  Lett.\  \textbf{118} (2017) 261803},
  \href{http://arxiv.org/abs/1702.06490}{{\normalfont\ttfamily
  arXiv:1702.06490}}\relax
\mciteBstWouldAddEndPuncttrue
\mciteSetBstMidEndSepPunct{\mcitedefaultmidpunct}
{\mcitedefaultendpunct}{\mcitedefaultseppunct}\relax
\EndOfBibitem
\bibitem{Nierste:2015zra}
U.~Nierste and S.~Schacht, \ifthenelse{\boolean{articletitles}}{\emph{{CP
  violation in $D^0\rightarrow \KS \KS$}},
  }{}\href{http://dx.doi.org/10.1103/PhysRevD.92.054036}{Phys.\ Rev.\
  \textbf{D92} (2015) 054036},
  \href{http://arxiv.org/abs/1508.00074}{{\normalfont\ttfamily
  arXiv:1508.00074}}\relax
\mciteBstWouldAddEndPuncttrue
\mciteSetBstMidEndSepPunct{\mcitedefaultmidpunct}
{\mcitedefaultendpunct}{\mcitedefaultseppunct}\relax
\EndOfBibitem
\bibitem{Brod:2011re}
J.~Brod, A.~L. Kagan, and J.~Zupan,
  \ifthenelse{\boolean{articletitles}}{\emph{{Size of direct CP violation in
  singly Cabibbo-suppressed D decays}},
  }{}\href{http://dx.doi.org/10.1103/PhysRevD.86.014023}{Phys.\ Rev.\
  \textbf{D86} (2012) 014023},
  \href{http://arxiv.org/abs/1111.5000}{{\normalfont\ttfamily
  arXiv:1111.5000}}\relax
\mciteBstWouldAddEndPuncttrue
\mciteSetBstMidEndSepPunct{\mcitedefaultmidpunct}
{\mcitedefaultendpunct}{\mcitedefaultseppunct}\relax
\EndOfBibitem
\bibitem{Dash:2017heu}
Belle collaboration, N.~Dash {\em et~al.},
  \ifthenelse{\boolean{articletitles}}{\emph{{Search for \CP violation and
  measurement of the branching fraction in the decay $\Dz \to \KS \KS$}},
  }{}\href{http://dx.doi.org/10.1103/PhysRevLett.119.171801}{Phys.\ Rev.\
  Lett.\  \textbf{119} (2017) 171801},
  \href{http://arxiv.org/abs/1705.05966}{{\normalfont\ttfamily
  arXiv:1705.05966}}\relax
\mciteBstWouldAddEndPuncttrue
\mciteSetBstMidEndSepPunct{\mcitedefaultmidpunct}
{\mcitedefaultendpunct}{\mcitedefaultseppunct}\relax
\EndOfBibitem
\bibitem{LHCb-PAPER-2015-030}
LHCb collaboration, R.~Aaij {\em et~al.},
  \ifthenelse{\boolean{articletitles}}{\emph{{Measurement of the
  time-integrated $\CP$ asymmetry in $\Dz\to\KS\KS$ decays}},
  }{}\href{http://dx.doi.org/10.1007/JHEP10(2015)055}{JHEP \textbf{10} (2015)
  055}, \href{http://arxiv.org/abs/1508.06087}{{\normalfont\ttfamily
  arXiv:1508.06087}}\relax
\mciteBstWouldAddEndPuncttrue
\mciteSetBstMidEndSepPunct{\mcitedefaultmidpunct}
{\mcitedefaultendpunct}{\mcitedefaultseppunct}\relax
\EndOfBibitem
\bibitem{Bonvicini:2000qm}
CLEO collaboration, G.~Bonvicini {\em et~al.},
  \ifthenelse{\boolean{articletitles}}{\emph{{Search for CP violation in $D^0
  \to \KS \pi^0$ and $D^0 \to \pi^0 \pi^0$ and $D^0 \to \KS \KS$ decays}},
  }{}\href{http://dx.doi.org/10.1103/PhysRevD.63.071101}{Phys.\ Rev.\
  \textbf{D63} (2001) 071101},
  \href{http://arxiv.org/abs/hep-ex/0012054}{{\normalfont\ttfamily
  arXiv:hep-ex/0012054}}\relax
\mciteBstWouldAddEndPuncttrue
\mciteSetBstMidEndSepPunct{\mcitedefaultmidpunct}
{\mcitedefaultendpunct}{\mcitedefaultseppunct}\relax
\EndOfBibitem
\bibitem{PDG2017}
Particle Data Group, C.~Patrignani {\em et~al.},
  \ifthenelse{\boolean{articletitles}}{\emph{{\href{http://pdg.lbl.gov/}{Review
  of particle physics}}},
  }{}\href{http://dx.doi.org/10.1088/1674-1137/40/10/100001}{Chin.\ Phys.\
  \textbf{C40} (2016) 100001}, and {\href{http://pdglive.lbl.gov/}{2017
  update}}\relax
\mciteBstWouldAddEndPuncttrue
\mciteSetBstMidEndSepPunct{\mcitedefaultmidpunct}
{\mcitedefaultendpunct}{\mcitedefaultseppunct}\relax
\EndOfBibitem
\bibitem{Alves:2008zz}
LHCb collaboration, A.~A. Alves~Jr.\ {\em et~al.},
  \ifthenelse{\boolean{articletitles}}{\emph{{The \lhcb detector at the LHC}},
  }{}\href{http://dx.doi.org/10.1088/1748-0221/3/08/S08005}{JINST \textbf{3}
  (2008) S08005}\relax
\mciteBstWouldAddEndPuncttrue
\mciteSetBstMidEndSepPunct{\mcitedefaultmidpunct}
{\mcitedefaultendpunct}{\mcitedefaultseppunct}\relax
\EndOfBibitem
\bibitem{LHCb-DP-2014-002}
LHCb collaboration, R.~Aaij {\em et~al.},
  \ifthenelse{\boolean{articletitles}}{\emph{{LHCb detector performance}},
  }{}\href{http://dx.doi.org/10.1142/S0217751X15300227}{Int.\ J.\ Mod.\ Phys.\
  \textbf{A30} (2015) 1530022},
  \href{http://arxiv.org/abs/1412.6352}{{\normalfont\ttfamily
  arXiv:1412.6352}}\relax
\mciteBstWouldAddEndPuncttrue
\mciteSetBstMidEndSepPunct{\mcitedefaultmidpunct}
{\mcitedefaultendpunct}{\mcitedefaultseppunct}\relax
\EndOfBibitem
\bibitem{Sjostrand:2006za}
T.~Sj\"{o}strand, S.~Mrenna, and P.~Skands,
  \ifthenelse{\boolean{articletitles}}{\emph{{PYTHIA 6.4 physics and manual}},
  }{}\href{http://dx.doi.org/10.1088/1126-6708/2006/05/026}{JHEP \textbf{05}
  (2006) 026}, \href{http://arxiv.org/abs/hep-ph/0603175}{{\normalfont\ttfamily
  arXiv:hep-ph/0603175}}\relax
\mciteBstWouldAddEndPuncttrue
\mciteSetBstMidEndSepPunct{\mcitedefaultmidpunct}
{\mcitedefaultendpunct}{\mcitedefaultseppunct}\relax
\EndOfBibitem
\bibitem{Sjostrand:2007gs}
T.~Sj\"{o}strand, S.~Mrenna, and P.~Skands,
  \ifthenelse{\boolean{articletitles}}{\emph{{A brief introduction to PYTHIA
  8.1}}, }{}\href{http://dx.doi.org/10.1016/j.cpc.2008.01.036}{Comput.\ Phys.\
  Commun.\  \textbf{178} (2008) 852},
  \href{http://arxiv.org/abs/0710.3820}{{\normalfont\ttfamily
  arXiv:0710.3820}}\relax
\mciteBstWouldAddEndPuncttrue
\mciteSetBstMidEndSepPunct{\mcitedefaultmidpunct}
{\mcitedefaultendpunct}{\mcitedefaultseppunct}\relax
\EndOfBibitem
\bibitem{LHCb-PROC-2010-056}
I.~Belyaev {\em et~al.}, \ifthenelse{\boolean{articletitles}}{\emph{{Handling
  of the generation of primary events in Gauss, the LHCb simulation
  framework}}, }{}\href{http://dx.doi.org/10.1088/1742-6596/331/3/032047}{{J.\
  Phys.\ Conf.\ Ser.\ } \textbf{331} (2011) 032047}\relax
\mciteBstWouldAddEndPuncttrue
\mciteSetBstMidEndSepPunct{\mcitedefaultmidpunct}
{\mcitedefaultendpunct}{\mcitedefaultseppunct}\relax
\EndOfBibitem
\bibitem{Lange:2001uf}
D.~J. Lange, \ifthenelse{\boolean{articletitles}}{\emph{{The EvtGen particle
  decay simulation package}},
  }{}\href{http://dx.doi.org/10.1016/S0168-9002(01)00089-4}{Nucl.\ Instrum.\
  Meth.\  \textbf{A462} (2001) 152}\relax
\mciteBstWouldAddEndPuncttrue
\mciteSetBstMidEndSepPunct{\mcitedefaultmidpunct}
{\mcitedefaultendpunct}{\mcitedefaultseppunct}\relax
\EndOfBibitem
\bibitem{Golonka:2005pn}
P.~Golonka and Z.~Was, \ifthenelse{\boolean{articletitles}}{\emph{{PHOTOS Monte
  Carlo: A precision tool for QED corrections in $Z$ and $W$ decays}},
  }{}\href{http://dx.doi.org/10.1140/epjc/s2005-02396-4}{Eur.\ Phys.\ J.\
  \textbf{C45} (2006) 97},
  \href{http://arxiv.org/abs/hep-ph/0506026}{{\normalfont\ttfamily
  arXiv:hep-ph/0506026}}\relax
\mciteBstWouldAddEndPuncttrue
\mciteSetBstMidEndSepPunct{\mcitedefaultmidpunct}
{\mcitedefaultendpunct}{\mcitedefaultseppunct}\relax
\EndOfBibitem
\bibitem{Allison:2006ve}
Geant4 collaboration, J.~Allison {\em et~al.},
  \ifthenelse{\boolean{articletitles}}{\emph{{Geant4 developments and
  applications}}, }{}\href{http://dx.doi.org/10.1109/TNS.2006.869826}{IEEE
  Trans.\ Nucl.\ Sci.\  \textbf{53} (2006) 270}\relax
\mciteBstWouldAddEndPuncttrue
\mciteSetBstMidEndSepPunct{\mcitedefaultmidpunct}
{\mcitedefaultendpunct}{\mcitedefaultseppunct}\relax
\EndOfBibitem
\bibitem{Agostinelli:2002hh}
Geant4 collaboration, S.~Agostinelli {\em et~al.},
  \ifthenelse{\boolean{articletitles}}{\emph{{Geant4: A simulation toolkit}},
  }{}\href{http://dx.doi.org/10.1016/S0168-9002(03)01368-8}{Nucl.\ Instrum.\
  Meth.\  \textbf{A506} (2003) 250}\relax
\mciteBstWouldAddEndPuncttrue
\mciteSetBstMidEndSepPunct{\mcitedefaultmidpunct}
{\mcitedefaultendpunct}{\mcitedefaultseppunct}\relax
\EndOfBibitem
\bibitem{LHCb-PROC-2011-006}
M.~Clemencic {\em et~al.}, \ifthenelse{\boolean{articletitles}}{\emph{{The
  \lhcb simulation application, Gauss: Design, evolution and experience}},
  }{}\href{http://dx.doi.org/10.1088/1742-6596/331/3/032023}{{J.\ Phys.\ Conf.\
  Ser.\ } \textbf{331} (2011) 032023}\relax
\mciteBstWouldAddEndPuncttrue
\mciteSetBstMidEndSepPunct{\mcitedefaultmidpunct}
{\mcitedefaultendpunct}{\mcitedefaultseppunct}\relax
\EndOfBibitem
\bibitem{Narsky:2014fya}
I.~Narsky and F.~C. Porter, {\em {Statistical analysis techniques in particle
  physics}}, Wiley-VCH, Weinheim, Germany, 2014\relax
\mciteBstWouldAddEndPuncttrue
\mciteSetBstMidEndSepPunct{\mcitedefaultmidpunct}
{\mcitedefaultendpunct}{\mcitedefaultseppunct}\relax
\EndOfBibitem
\bibitem{Johnson:1949zj}
N.~L. Johnson, \ifthenelse{\boolean{articletitles}}{\emph{{Systems of frequency
  curves generated by methods of translation}},
  }{}\href{http://dx.doi.org/10.1093/biomet/36.1-2.149}{Biometrika \textbf{36}
  (1949) 149}\relax
\mciteBstWouldAddEndPuncttrue
\mciteSetBstMidEndSepPunct{\mcitedefaultmidpunct}
{\mcitedefaultendpunct}{\mcitedefaultseppunct}\relax
\EndOfBibitem
\bibitem{LHCb:2012fb}
LHCb collaboration, R.~Aaij {\em et~al.},
  \ifthenelse{\boolean{articletitles}}{\emph{{Measurement of the $D^{\pm}$
  production asymmetry in 7 TeV pp collisions}},
  }{}\href{http://dx.doi.org/10.1016/j.physletb.2012.11.038}{Phys.\ Lett.\
  \textbf{B718} (2013) 902},
  \href{http://arxiv.org/abs/1210.4112}{{\normalfont\ttfamily
  arXiv:1210.4112}}\relax
\mciteBstWouldAddEndPuncttrue
\mciteSetBstMidEndSepPunct{\mcitedefaultmidpunct}
{\mcitedefaultendpunct}{\mcitedefaultseppunct}\relax
\EndOfBibitem
\bibitem{LHCb-PAPER-2014-042}
LHCb collaboration, R.~Aaij {\em et~al.},
  \ifthenelse{\boolean{articletitles}}{\emph{{Measurement of the $\Bzb$--$\Bz$
  and $\Bsb$--$\Bs$ production asymmetries in $\proton\proton$ collisions at
  $\sqrt{s}=7$\tev}},
  }{}\href{http://dx.doi.org/10.1016/j.physletb.2014.10.005}{Phys.\ Lett.\
  \textbf{B739} (2014) 218},
  \href{http://arxiv.org/abs/1408.0275}{{\normalfont\ttfamily
  arXiv:1408.0275}}\relax
\mciteBstWouldAddEndPuncttrue
\mciteSetBstMidEndSepPunct{\mcitedefaultmidpunct}
{\mcitedefaultendpunct}{\mcitedefaultseppunct}\relax
\EndOfBibitem
\bibitem{Aaij:2017mso}
LHCb collaboration, R.~Aaij {\em et~al.},
  \ifthenelse{\boolean{articletitles}}{\emph{{Measurement of $B^0$, $B^0_s$,
  $B^+$ and $\Lambda^0_b$ production asymmetries in 7 and 8 TeV proton-proton
  collisions}},
  }{}\href{http://dx.doi.org/10.1016/j.physletb.2017.09.023}{Phys.\ Lett.\
  \textbf{B774} (2017) 139},
  \href{http://arxiv.org/abs/1703.08464}{{\normalfont\ttfamily
  arXiv:1703.08464}}\relax
\mciteBstWouldAddEndPuncttrue
\mciteSetBstMidEndSepPunct{\mcitedefaultmidpunct}
{\mcitedefaultendpunct}{\mcitedefaultseppunct}\relax
\EndOfBibitem
\bibitem{Enz}
C.~P. Enz and R.~R. Lewis, \ifthenelse{\boolean{articletitles}}{\emph{{On the
  phenomenological description of CP violation for K mesons and its
  consequences}}, }{}Helv.\ Phys.\ Acta \textbf{38} (1965) 860\relax
\mciteBstWouldAddEndPuncttrue
\mciteSetBstMidEndSepPunct{\mcitedefaultmidpunct}
{\mcitedefaultendpunct}{\mcitedefaultseppunct}\relax
\EndOfBibitem
\end{mcitethebibliography}

\clearpage
\centerline{\large\bf LHCb collaboration}
\begin{flushleft}
\small
R.~Aaij$^{27}$,
B.~Adeva$^{41}$,
M.~Adinolfi$^{48}$,
C.A.~Aidala$^{73}$,
Z.~Ajaltouni$^{5}$,
S.~Akar$^{59}$,
P.~Albicocco$^{18}$,
J.~Albrecht$^{10}$,
F.~Alessio$^{42}$,
M.~Alexander$^{53}$
% ...

{\footnotesize \it
% ...
$ ^{5}$Clermont Universit{\'e}, Universit{\'e} Blaise Pascal, CNRS/IN2P3, LPC, Clermont-Ferrand, France\\
% ...
$ ^{10}$Fakult{\"a}t Physik, Technische Universit{\"a}t Dortmund, Dortmund, Germany\\
% ...
$ ^{18}$INFN Laboratori Nazionali di Frascati, Frascati, Italy\\
% ...
$ ^{27}$Nikhef National Institute for Subatomic Physics, Amsterdam, Netherlands\\
% ...
$ ^{41}$Instituto Galego de F{\'\i}sica de Altas Enerx{\'\i}as (IGFAE), Universidade de Santiago de Compostela, Santiago de Compostela, Spain\\
$ ^{42}$European Organization for Nuclear Research (CERN), Geneva, Switzerland\\
% ...
$ ^{48}$H.H. Wills Physics Laboratory, University of Bristol, Bristol, United Kingdom\\
% ...
$ ^{53}$School of Physics and Astronomy, University of Glasgow, Glasgow, United Kingdom\\
% ...
$ ^{59}$University of Cincinnati, Cincinnati, OH, United States\\
% ...
$ ^{61}$Syracuse University, Syracuse, NY, United States\\
% ...
$ ^{73}$University of Michigan, Ann Arbor, United States, associated to $^{61}$\\
% ...
}
\end{flushleft}

\end{document}